%% file: FourTaus.tex
\documentclass[a4paper,12pt]{article}
\usepackage{a4,epsfig,amssymb,rotating, subfigure}
\usepackage{slashed}
\usepackage{multirow}
\def\ra{\rightarrow}

\newcommand{\PZz}{\mathrm{Z}}

\newcommand{\GEV}{\mbox{$\mathrm{GeV}$}}

\newcommand{\GEVcc}{\mbox{$\mathrm{GeV}/{{\it c}^2}$}}

\newcommand{\qqbar}{\mbox{$\mathrm{q\bar{q}}$}}

\newcommand{\bbbar}{\mbox{$\mathrm{b\bar{b}}$}}

\newcommand{\ma}{m_\A}
\newcommand{\mh}{m_\mathrm{h}}
\newcommand{\eplus}{\mathrm{e}^+}
\newcommand{\eminus}{\mathrm{e}^-}
\newcommand{\mZ}{m_{\PZz}}
\newcommand{\A}{\mathrm{a}}
\newcommand{\jtrk}{\ensuremath{n^\mathrm{track}}}
\font\ninerm=cmr9

\begin{document}

\begin{titlepage}

{\Large\centerline{EUROPEAN ORGANISATION FOR NUCLEAR RESEARCH (CERN)}}
\vspace{1cm}

\begin{center}
\vspace{3cm}
\boldmath
{\huge\bf Search for neutral Higgs bosons decaying into four taus at LEP2}



{\Large \vspace{7ex} The ALEPH Collaboration$^*$)}
\unboldmath
\end{center}

\vspace{1cm}
\begin{center}
{\bf Abstract}
\end{center}
A search for the production and non-standard decay of a Higgs boson, $\mathrm{h}$, into four taus through intermediate pseudoscalars, $\A $, is conducted on 683 pb$^{-1}$ of data collected by the ALEPH experiment at centre-of-mass energies from 183 to 209~\GEV. No excess of events above background is observed, and exclusion limits are placed on the combined production cross section times branching ratio, $\xi^2 = \frac{\sigma(\eplus \eminus\ra \PZz\mathrm{h})}{\sigma_{\rm SM}(\eplus \eminus\ra \PZz\mathrm{h})}\times B(\mathrm{h}\ra \A\A)\times B(\A\ra \tau^+\tau^-)^2$.  For $\mh< 107~\GEVcc$ and  $4<\ma<10~\GEVcc$, $\xi^2>1$ is excluded at the 95\% confidence level. \\[1.5cm]
\vspace{1cm}
\vfill
\centerline{\it Submitted to the Journal of High Energy Physics (JHEP)}
\vskip .5cm
\noindent
--------------------------------------------\hfil\break
{\ninerm $^*$ See next pages for the list of authors}

\end{titlepage}
\newpage

\input{authb.tex}

\section{Introduction}

Searches conducted at LEP2 have excluded the standard model (SM) Higgs boson decaying into $\bbbar$ or $\tau^+\tau^-$ for masses below $114.4~\GEVcc~$\cite{Barate:2003sz}.  The LEP experiments observed a small excess in the $\bbbar$ final state for a Higgs boson mass around 100~$\GEVcc$, which is consistent with background fluctuations or SM-like production with a reduced branching ratio into $\bbbar$~\cite{Schael:2006cr,Dermisek:2005gg}. This excess, the mild tension with electroweak precision tests~\cite{Barbieri:1999tm}, and the fine-tuning needed in the minimal supersymmetric standard model (MSSM) have prompted the consideration of models with exotic Higgs boson decays, such as those of the next-to-minimal supersymmetric standard model (NMSSM)~\cite{Dermisek:2005ar,Dermisek:2007yt} as well as more general frameworks~\cite{Chang:2005ht,Chang:2008cw}. In these models, new decay channels can dominate over $\mathrm{h} \ra \bbbar$ and render the Higgs boson  ``invisible'' for conventional searches. In particular, a Higgs boson decaying into two light pseudoscalars is well motivated by these models and results in a four-body final state as the pseudoscalars decay into light fermions. Scenarios with a dominant $\mathrm{h}\ra 2\A \ra 4\mathrm{b}$ decay are constrained by the LEP experiments for mass up to  $110~\GEVcc$~\cite{Schael:2006cr,Abbiendi:2004ww, Delphi:2007ge}. A general search for $\mathrm{h} \ra \A\A$ with $\A \ra \mathrm{gg},\mathrm{c}\bar{\mathrm{c}},\tau^+\tau^-$ was performed by the OPAL Collaboration~\cite{Abbiendi:2002in}, but the analysis was restricted to a Higgs boson mass, $\mh$, in the range 45--86$~\GEVcc$.   
A search for $\mathrm{h}\ra 2\A \ra 2\mu 2\tau$ was recently reported by the D0 Collaboration, which resulted in limits that are a factor of 1-4 larger than the expected production cross section assuming branching ratios of the pseudoscalar as predicted by the NMSSM~\cite{Abazov:2009yi}. In this paper, a search using ALEPH data is presented  for $\mathrm{h}\ra 2\A\ra 4\tau$ up to $\mh \approx 110~\GEVcc$.

\begin{figure}[hb]
\begin{center}
\vspace{-.1in}
\includegraphics[scale=.5]{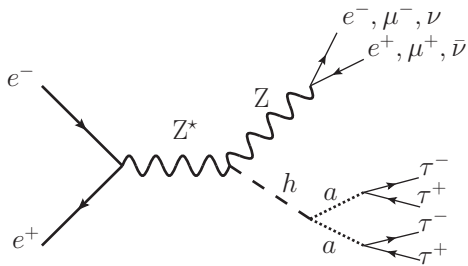}
\end{center}
\caption{Higgs boson production and decay modes considered in this analysis.}
\label{fig:topo}
\end{figure}

The pseudoscalar $\A$ may arise from a  two Higgs doublet model, as in the MSSM, or it can include a component from an additional singlet field as in the NMSSM. These possibilities differ in their details and relations between model parameters. The present search is performed in a model-independent manner and simply adopts the two main characteristics of the pseudoscalar: the coupling to a Higgs boson resulting in $\mathrm{h}\ra \A\A$ decay and the coupling to SM fermions proportional to their Yukawa couplings.  The present analysis concentrates on the region $2m_\tau< \ma \lesssim 2m_\mathrm{b}$, where the $\A\ra\tau^+\tau^-$ decay mode is expected to be substantial.  The Higgs boson production mode considered here is the Higgsstrahlung process, shown in Fig.~\ref{fig:topo} with $\PZz\ra \eplus\eminus,~\mu^+\mu^-,~ \nu\bar{\nu}$.

This paper is organized as follows. The ALEPH detector is described in Section \ref{sec:aleph}. Simulation of background and signal processes are described in Section \ref{sec:sim}. The event selection criteria are described in Section \ref{sec:events}. Systematic uncertainties are discussed in Section \ref{sec:systematics}. Finally, results  and conclusions are presented in Sections \ref{sec:results} and ~\ref{sec:conclusions}, respectively.

\section{The ALEPH Detector}
\label{sec:aleph}
A detailed description of the ALEPH detector can be found in
Ref.~\cite{aleph_det} and of its performance in Ref.~\cite{aleph-perf}. 
Charged particles are detected in the central part, which consists of 
a precision silicon vertex detector, a cylindrical drift chamber 
 and a large time projection chamber (TPC), together measuring 
up to 31 space points along the charged particle trajectories.
A 1.5\,T axial magnetic field is provided by a superconducting solenoidal 
coil. Charged particle transverse momenta are reconstructed with a 
1/$p_{T}$ resolution of 
$\left( 6\times 10^{-4} \bigoplus 5 \times 10^{-3}/p_{T} \right)~({\rm GeV}/c)^{-1}$.
The charged particle tracks used in the present analysis (and simply called
{\it tracks}) are reconstructed with at least four hits in the TPC, and 
originate from within a cylinder of length 20\,cm and radius 2\,cm coaxial 
with the beam and centred at the nominal collision point. 


Electrons and photons are identified by the characteristic longitudinal
and transverse development of the associated showers in the electromagnetic 
calorimeter, a 22-radiation-length thick sandwich of lead planes and 
proportional wire chambers with fine readout segmentation. A
relative energy resolution of $0.18/\sqrt{E}$ ($E$ in GeV) is achieved
for isolated electrons and photons.

Muons are identified by their characteristic penetration pattern in the 
hadron calorimeter, a 1.2\,m thick yoke interleaved with 23 layers of 
streamer tubes, together with two surrounding double-layers of muon chambers. 
In association with the electromagnetic calorimeter, the hadron calorimeter 
also provides a measurement of the hadronic energy with a relative resolution 
of $0.85/\sqrt{E}$ ($E$ in GeV).

The total visible energy is measured with an 
energy-flow reconstruction algorithm which combines all the above 
measurements~\cite{aleph-perf}.
The relative resolution on the total visible energy is $0.60/\sqrt{E}$ for 
high multiplicity final states. In addition to the total visible-energy 
measurement, the energy-flow reconstruction algorithm also provides 
a list of reconstructed objects, classified as charged particles, photons 
and neutral hadrons, and called {\it energy-flow objects} in the 
following. These energy-flow objects are the 
basic entities used in the present analysis.

Below polar angles of 12$^\circ$ and down to 34 mrad from the beam axis, 
the acceptance is closed at both ends of the experiment by the luminosity 
calorimeter (LCAL) and a tungsten-silicon calorimeter 
(SICAL) originally designed for the LEP\,1 luminosity measurement. 


The average centre-of-mass energies at which the machine operated and the corresponding integrated luminosities used in this analysis are presented in Table \ref{tbl:EcmLumi}.

\begin{table}[hb]
\begin{center}
\caption{Integrated luminosities collected at the different average centre-of-mass energies.}
\vspace{2pt}
\label{tbl:EcmLumi}
\begin{tabular}[c]{c|c|c|c|c|c|c|c|c}
$E_{\rm CM}(\GEV)$ & 182.65 & 188.63 & 191.58 & 195.52 & 199.52 & 201.62 & 204.86 & 206.53 \\ 
\hline
$ \int \mathcal{L}dt~(\textrm{pb}^{-1})$ & 56.8 & 174.2 & 28.9 & 79.9 & 86.3 &41.9 & 81.4 & 133.2\\
\end{tabular}
\end{center}
\end{table}

\section{Signal and Background Generation}
\label{sec:sim}

Both signal and background were generated for all centre-of-mass energies shown in Table~\ref{tbl:EcmLumi} using the \textsc{geant3}-based simulation of ALEPH~\cite{Geant3}. Backgrounds were generated with a variety of generators listed in Table~\ref{tbl:Background}. 
The $\gamma\gamma$ initiated and Bhabha samples are 10--30 times larger than the data sample while others are 300--1000 times larger.  The contribution to the $\mathrm{We}\nu$ process from electrons emitted close to the beam axis is not included in \textsc{koralw} and is generated with \textsc{pythia}.  The \textsc{hzha03} generator~\cite{HZHA:Janot} was used to generate 3000 signal events (with $\mathrm{h}\ra \A\A$ followed by $\A \ra \tau^+\tau^-$) for each of the three $\PZz$ decay channels considered and for each combination of Higgs boson and pseudoscalar masses in the ranges $70 < \mh < 114~\GEVcc$ and $4 < \ma < 12~\GEVcc$ in steps of $2~\GEVcc$.


\begin{table}[h]
\begin{center}
\caption{Details on SM background processes and their categorisation. Fragmentation, hadronisation and final state radiation were simulated with $\textsc{pythia 6.1}$~\cite{Sjostrand:2000wi}.  {\sc photos}~\cite{Barberio:1993qi} was used to model final state radiation,  and {\sc tauola}~\cite{Jadach:1993hs} was used for tau decays.
 More details can be found in Ref. \cite{Heister:2004wr}.}
\label{tbl:Background}
\vspace{5pt}
\begin{tabular}[c]{|c|c|c|c|}
\hline
Category & Process & Software \\
\hline
\multirow{5}{*}{2f}   &  $\eplus\eminus\ra \PZz/\gamma^* \ra \qqbar(\gamma)$ & $\textsc{kk 4.14} $~\cite{Jadach:1999vf} \\
		    &  Bhabha and $\eplus\eminus\ra \PZz/\gamma^* \ra \eplus\eminus(\gamma)$	& $\textsc{bhwide 1.01} $~\cite{Jadach:1995nk} \\
		    &  $\eplus\eminus\ra \PZz/\gamma^* \ra \mu^+\mu^-(\gamma)$  & $\textsc{kk 4.14} $~\cite{Jadach:1999vf} \\ 
		    & $\eplus\eminus\ra \PZz/\gamma^* \ra \tau^+\tau^-(\gamma)$  & $\textsc{kk 4.14} $~\cite{Jadach:1999vf} \\
			& $\eplus\eminus\ra \PZz\ra\nu\bar{\nu}(\gamma)$       & $\textsc{pythia 6.1} $~\cite{Sjostrand:2000wi}  \\ 
\hline
\multirow{5}{*}{4f}      & $\eplus\eminus\ra \PZz/\gamma^* \ra \mathrm{W^+W^-}$                & $\textsc{koralw 1.51} $~\cite{Jadach:2001mp} \\ 
		    & $\eplus\eminus\ra \PZz\PZz$                                & $\textsc{pythia 6.1} $~\cite{Sjostrand:2000wi} \\ 
		    & $\eplus\eminus \ra \PZz \, \eplus\eminus$                        & $\textsc{pythia 6.1} $~\cite{Sjostrand:2000wi} \\ 
		     & $\eplus\eminus\ra \PZz \, \nu\bar{\nu}$       & $\textsc{pythia 6.1} $~\cite{Sjostrand:2000wi}  \\ 
		    & $\eplus\eminus\ra \mathrm{W}^{\pm}\mathrm{e}^{\mp} \nu$       & $\textsc{pythia 6.1} $~\cite{Sjostrand:2000wi}  \\ 

\hline
\multirow{2}{*}{$\gamma\gamma$}    & $\gamma\gamma \ra \ell^+\ell^-$                     & $\textsc{phot02} $~\cite{phot02,phot02:aleph} \\ 
 	             & $\gamma\gamma \ra \qqbar$                         & $\textsc{phot02} $~\cite{phot02,phot02:aleph} \\ 
\hline
\multirow{1}{*}{n$\gamma$}     
		       & $\eplus\eminus\ra n\gamma$                          & $\textsc{pythia 6.1} $~\cite{Sjostrand:2000wi}  \\ 
\hline
\end{tabular}

\end{center}
\end{table}


\section{Event Selection}
\label{sec:events}
A detailed description of the event selection criteria for the different $\PZz$ decays, namely $\PZz\ra \eplus\eminus$, $\PZz\ra \mu^+\mu^-$, and $\PZz\ra \nu\bar{\nu}$ is presented below. The event topology is discussed first, followed by an explanation of how the visible decay products of the taus were treated. The $\PZz$ reconstruction algorithm is then described and, finally, a detailed list of the cuts employed in the analysis is given. 


For the mass range considered, the Higgs boson is produced approximately at rest, and thus the decay $\mathrm{h}\ra 2\A\ra 4\tau$ results in a pair of taus recoiling against another pair of taus. The $\textsc{jade}$ algorithm~\cite{Bartel:1986mf,Bethke:1988zc}  was employed to cluster into jets all energy-flow objects except for those identified as energetic, isolated photons, energy deposits in the $\textsc{lcal}$ and $\textsc{sical}$, and the two hardest, oppositely-charged leptons in the case of the $\PZz\ra \ell^+\ell^-$ channels. Given that each jet is expected to arise from the on-shell decay $\A\ra \tau^+\tau^-$, an effective way to target the signal topology is to use the $\textsc{jade}$ algorithm with $y_{\rm cut}$ chosen to merge proto-jets up to a mass of $m_{\rm jet} = 15~\GEVcc$. This technique allows for an accurate clustering of the two tau pairs separately, thus rendering individual tau identification unnecessary and making track-multiplicity an effective discriminator against background.
 
Because the taus from the same $\A$ decay are highly collimated, the identification of jets containing the decay products of two taus was based only on the track multiplicity of the jets, denoted $\jtrk_i$, with the index $i$ ordered in decreasing jet energy.  Because the tau predominantly decays either to one charged particle (``one-prong'' decay) or three charged particles (``three-prong'' decay), each jet is expected to contain two, four, or six tracks. To maximize the tracking efficiency, the jets were required to be well contained in the tracking volume.  No distinction was made between leptonic and hadronic decays.  

The $\PZz\ra\ell^+\ell^-$ decay is often accompanied by additional photons from final state radiation, which can carry substantial momentum.  An object is considered as an isolated photon if it is identified as a photon by the energy-flow algorithm, has $E>10~\GEV$, and contains less than $5\%$ of the visible energy of the event in a cone of $10^\circ$ around it.  The photon was considered part of the candidate $\PZz$ system when the  invariant mass of the $\ell^+\ell^-\gamma$ system was closer to the $\PZz$ mass than the invariant mass of the lepton pair alone.  
This algorithm resulted in an increase of $\sim 20\%$ in the signal efficiency after the $\PZz$ mass window cut, $80 < \mZ < 102~\GEVcc$. 

For each of the channels below, a loose selection and final selection are presented.  The {loose selection} isolates the broad characteristics of the signal events and allows for comparison of the data and simulated backgrounds.

\subsection{$\PZz\ra \ell^+\ell^-$}

The loose selection consisted of the following requirements.  An $\eplus\eminus$ or $\mu^+\mu^-$ pair and the presence of two jets (or 3 jets with $\jtrk_3 \le 2$ ) were required for consistency with the final state of the signal.  The three-jet events are kept to recover signal efficiency for events with converted photon arising from final state radiation.   Proper containment of the jet in the tracking volume was ensured by requiring $|\cos\theta_{\rm j_1}| < 0.9$ and $|\cos\theta_{\rm j_2}| < 0.9$, where $\theta_{{\rm j}_i}$ is the angle of the $i^{\rm th}$ jet with respect to the beam axis .  Additional lepton isolation was imposed by requiring that  a cone of $10^\circ$ around each lepton contained less than $5\%$ of the visible energy of the event and $\cos\theta_{\rm jl}^{\rm min}<0.95$, where $\theta_{\rm jl}^{\rm min}$ is the minimum angle between each pairing of a jet and lepton.  

\begin{figure}[h]
\begin{center}
\subfigure[]{\includegraphics[scale=.41]{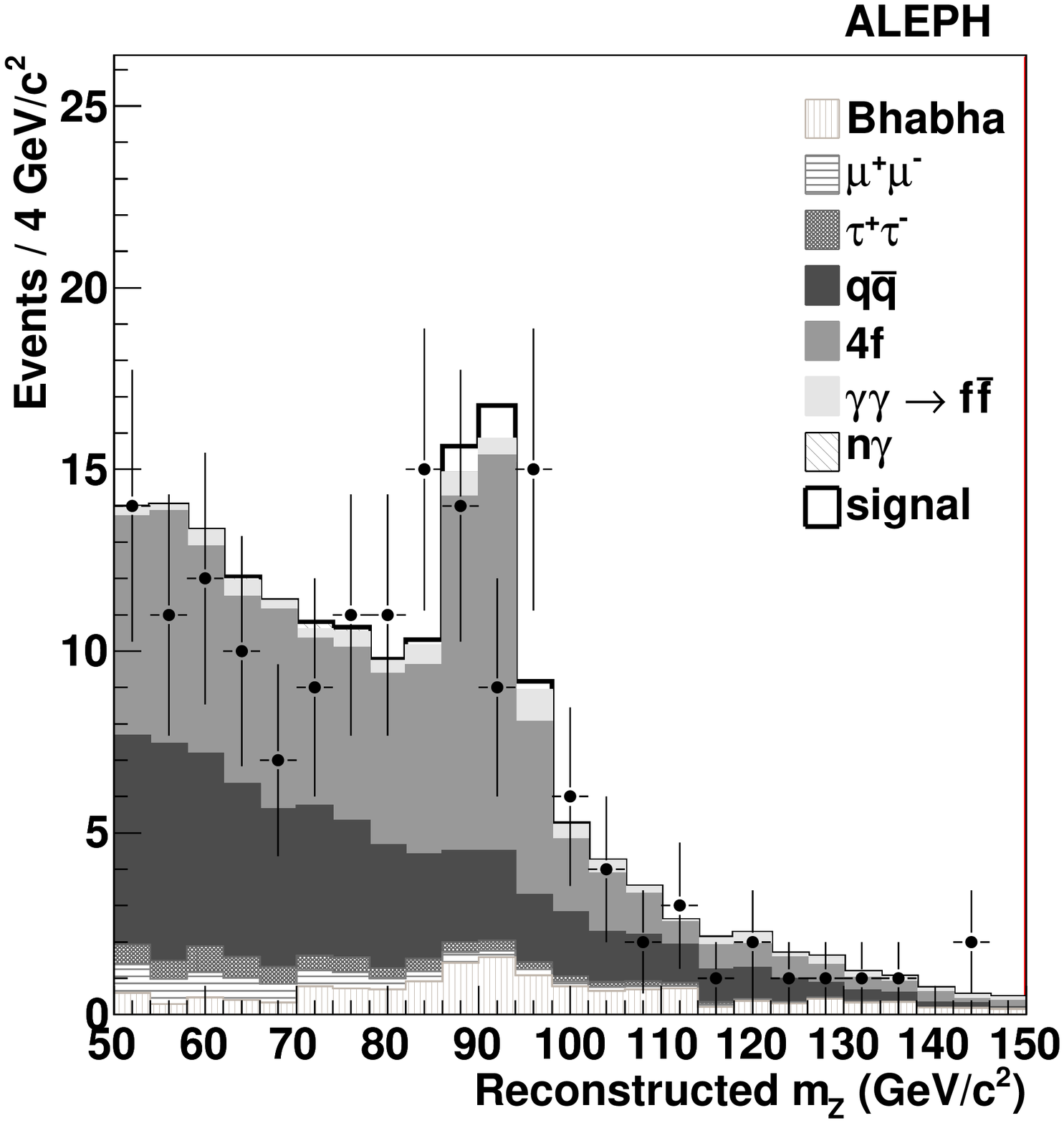}}%
\subfigure[]{\includegraphics[scale=.41]{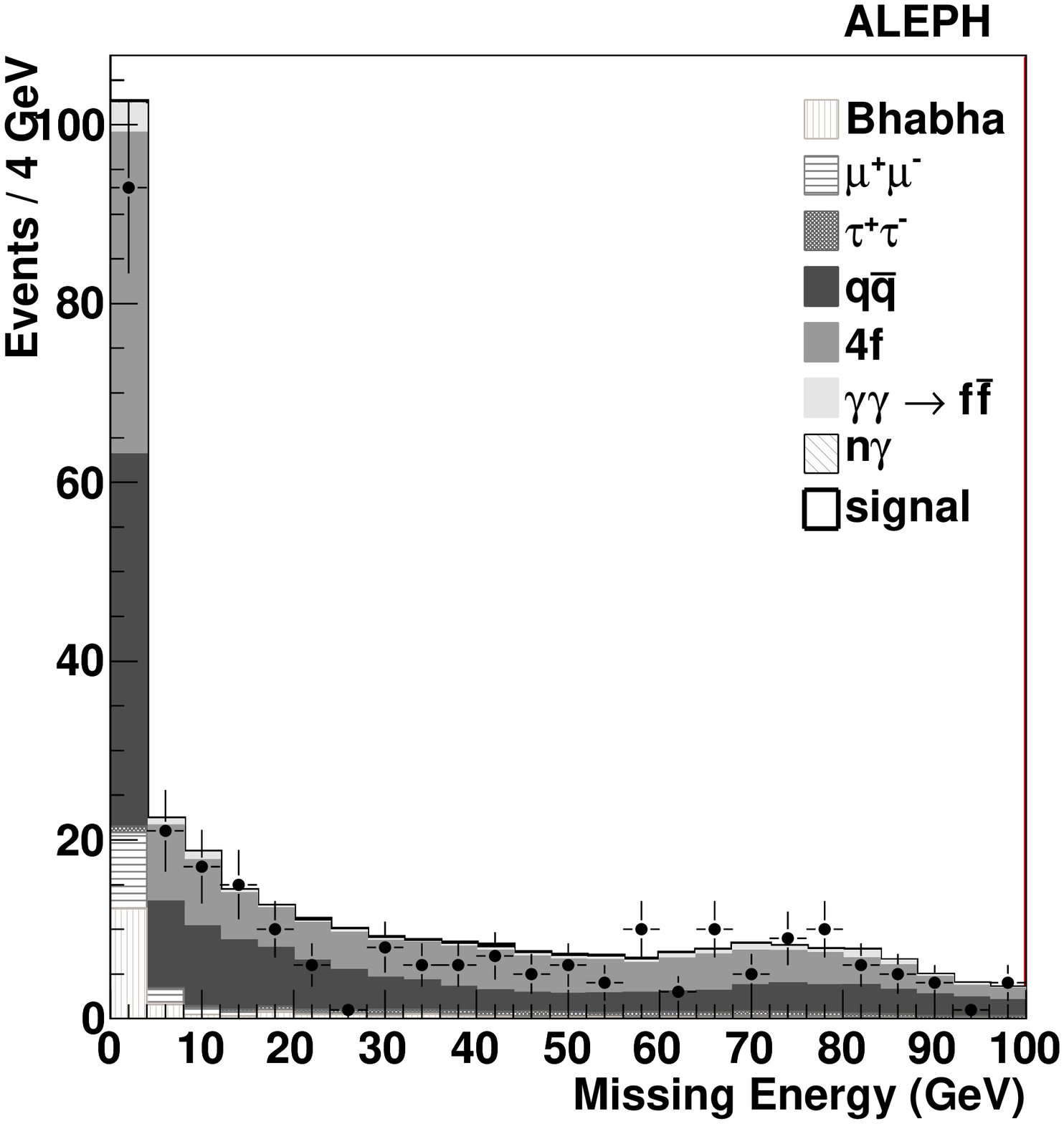}}\\
\subfigure[]{\includegraphics[scale=.41]{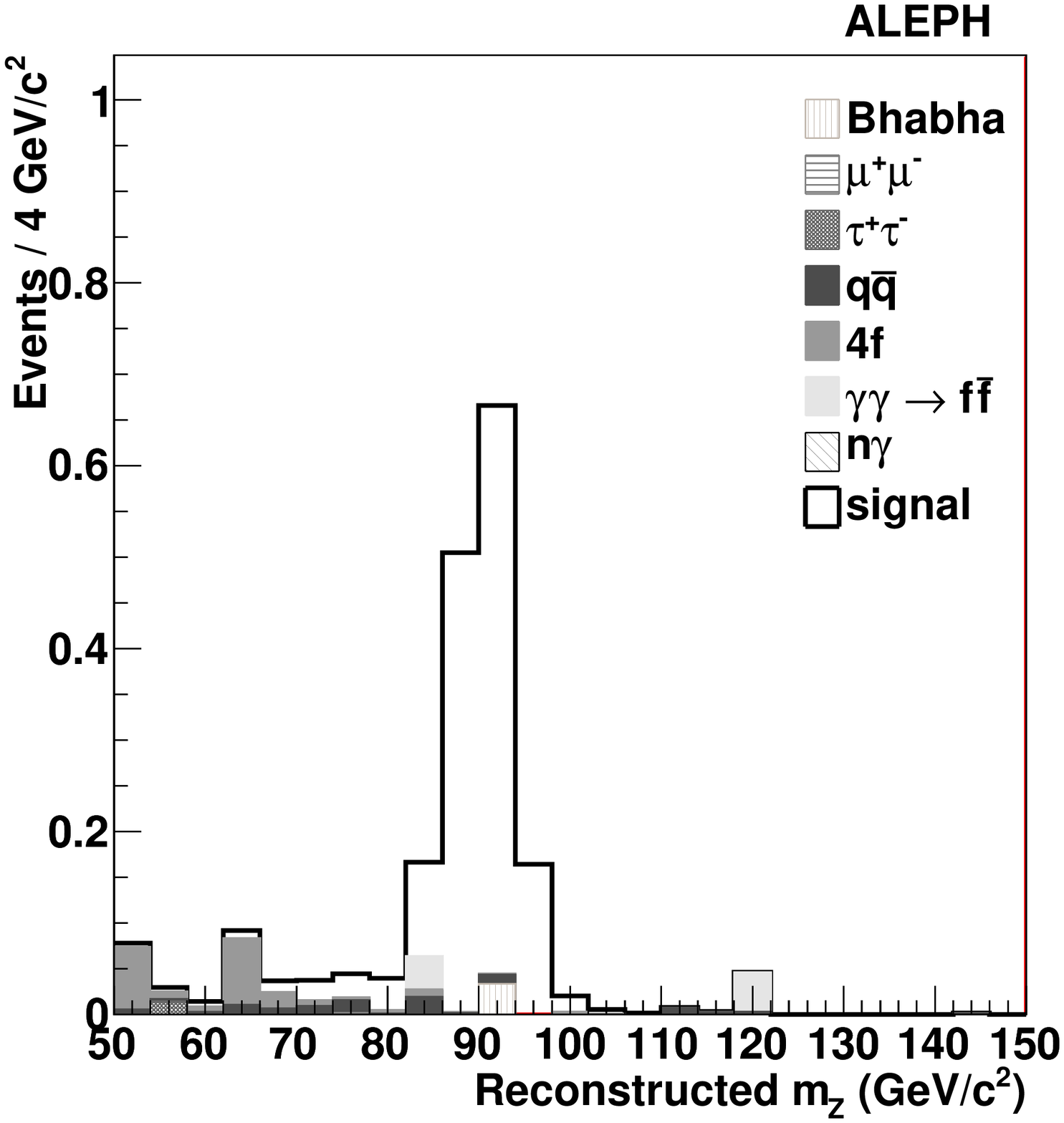}}%
\subfigure[]{\includegraphics[scale=.41]{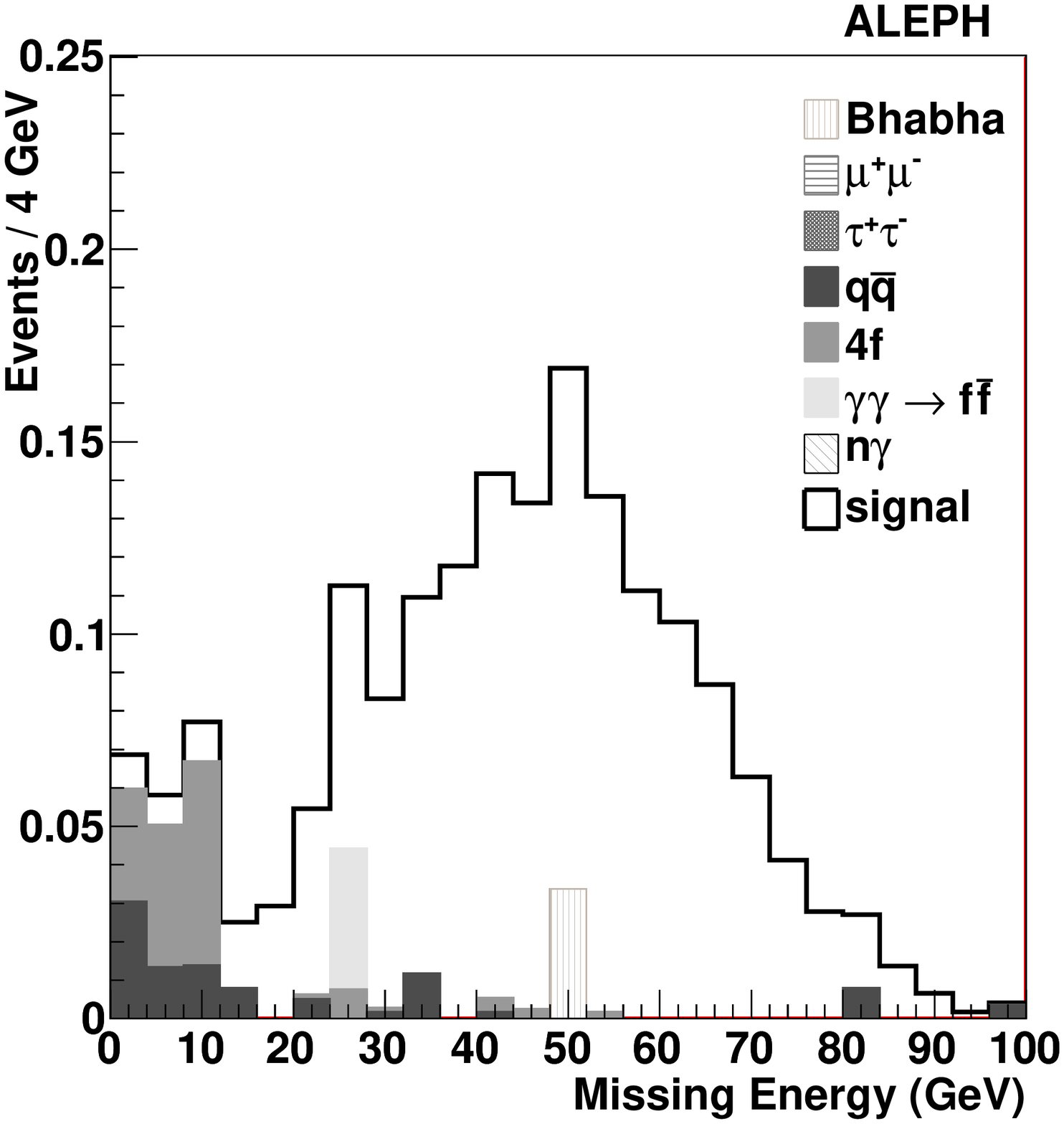}}
\end{center}
\caption{Distributions for the $\PZz\ra \eplus\eminus$ channel after the loose selection for (a) the reconstructed $\PZz$ invariant mass  and (b) missing energy, where signal corresponds to $\mh = 100~\GEVcc$, $\ma = 4~\GEVcc$ with $\xi^2=1$ (see text).  The same distributions are shown in (c) and (d) after the final selection, excluding any requirements on the variable shown.}
\label{fig:ElectronPlots}
\end{figure}

\begin{figure}[h]
\begin{center}
\subfigure[]{\includegraphics[scale=.41]{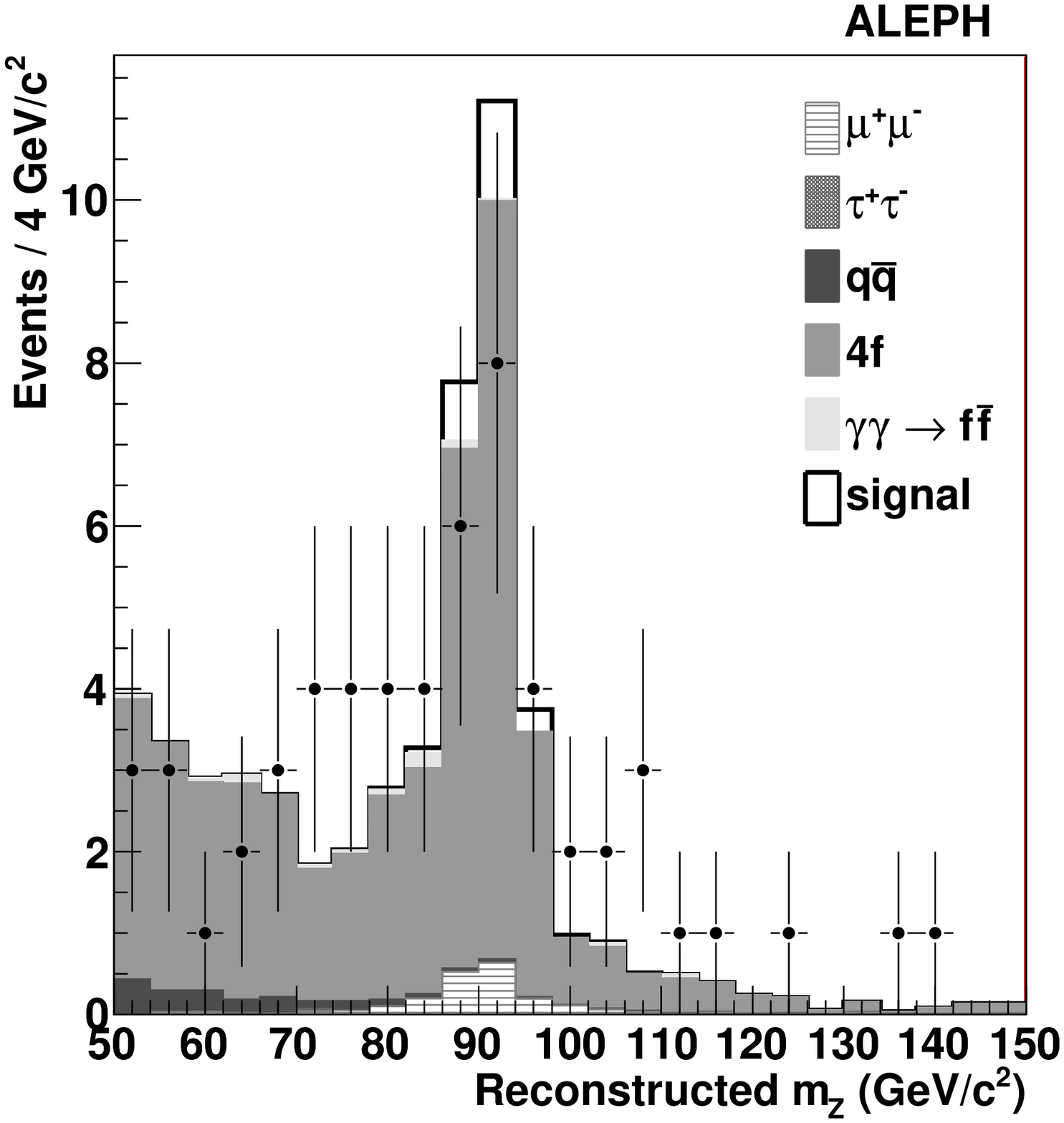}}%
\subfigure[]{\includegraphics[scale=.41]{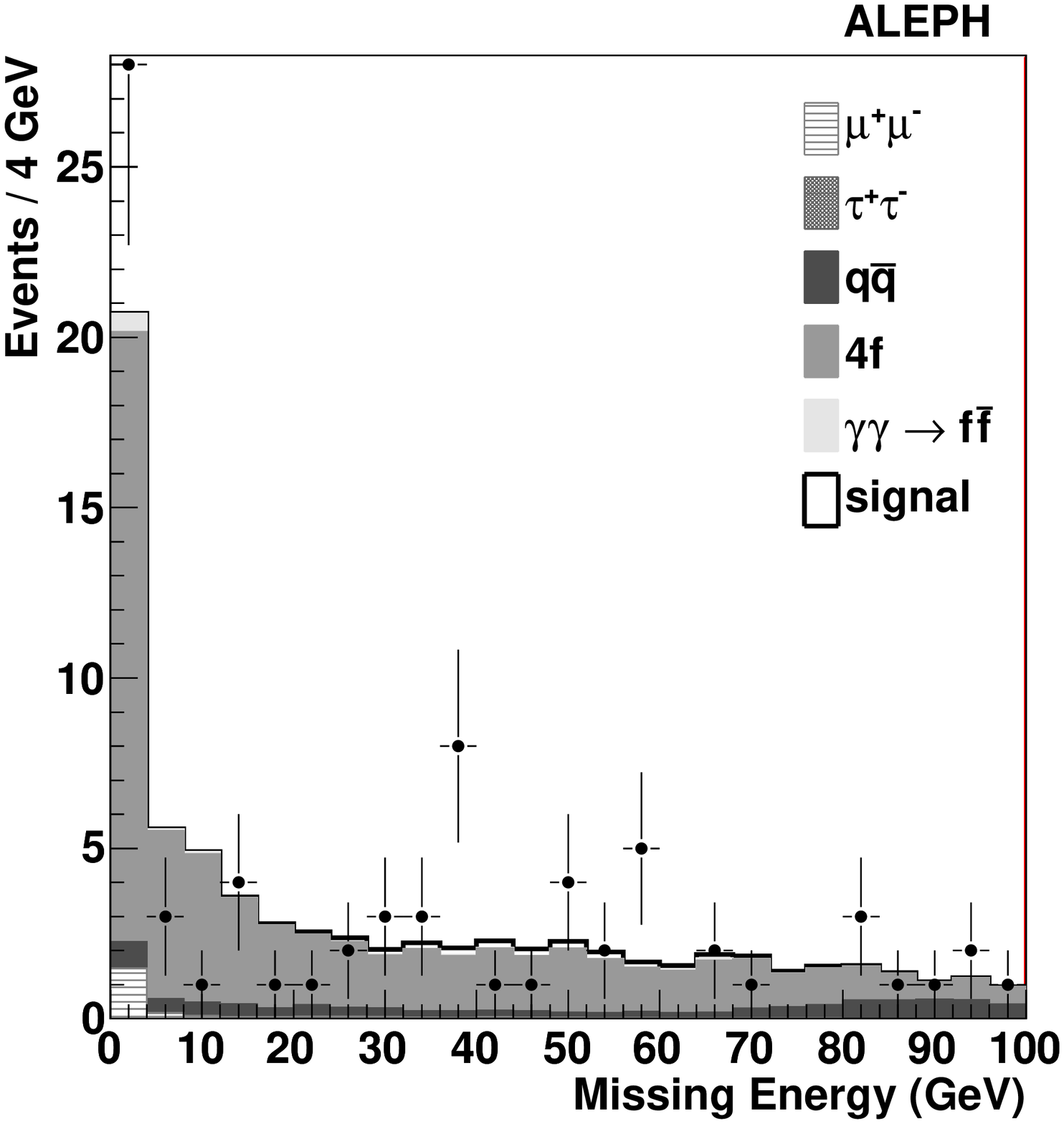}}\\
\subfigure[]{\includegraphics[scale=.41]{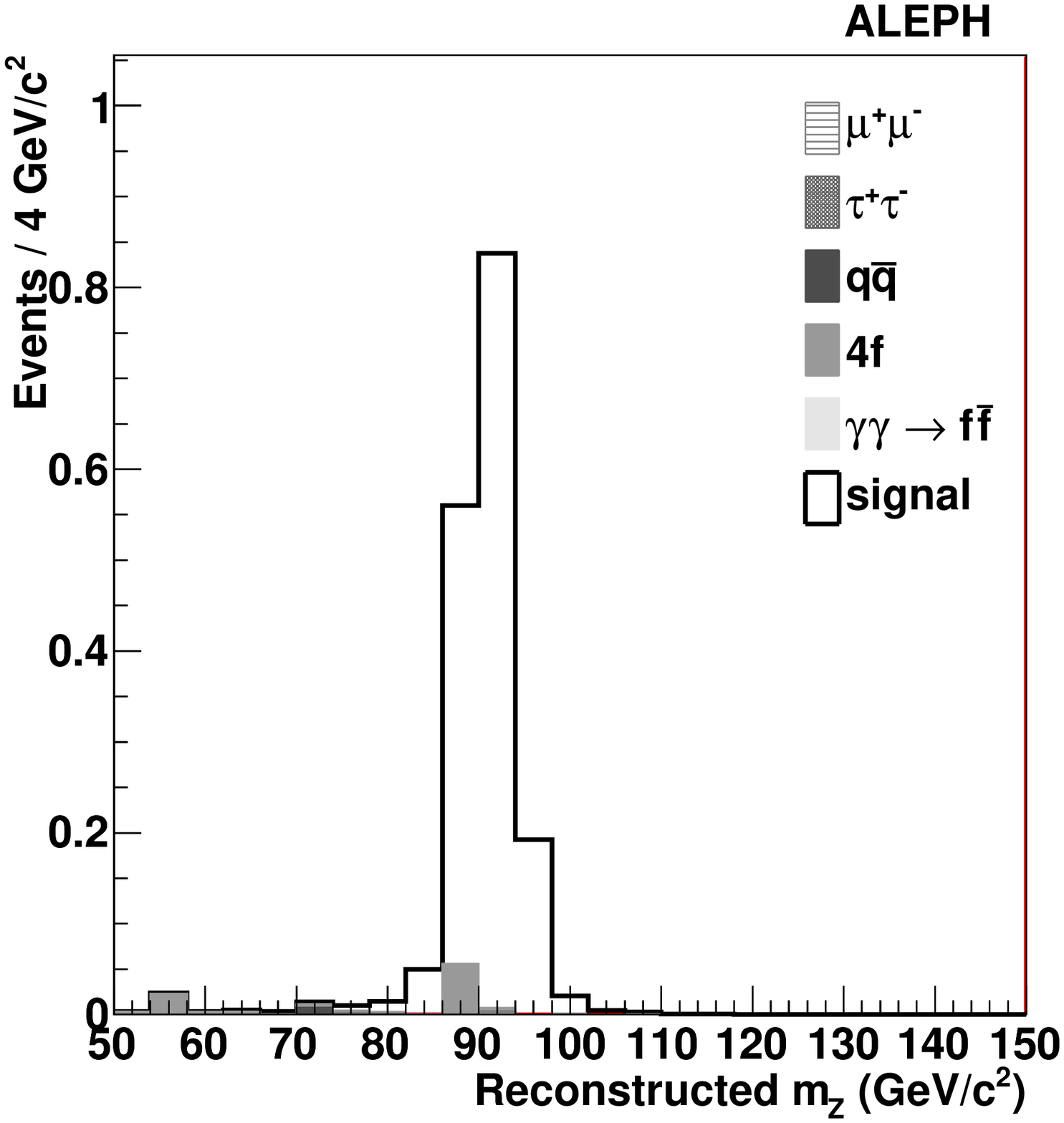}}%
\subfigure[]{\includegraphics[scale=.41]{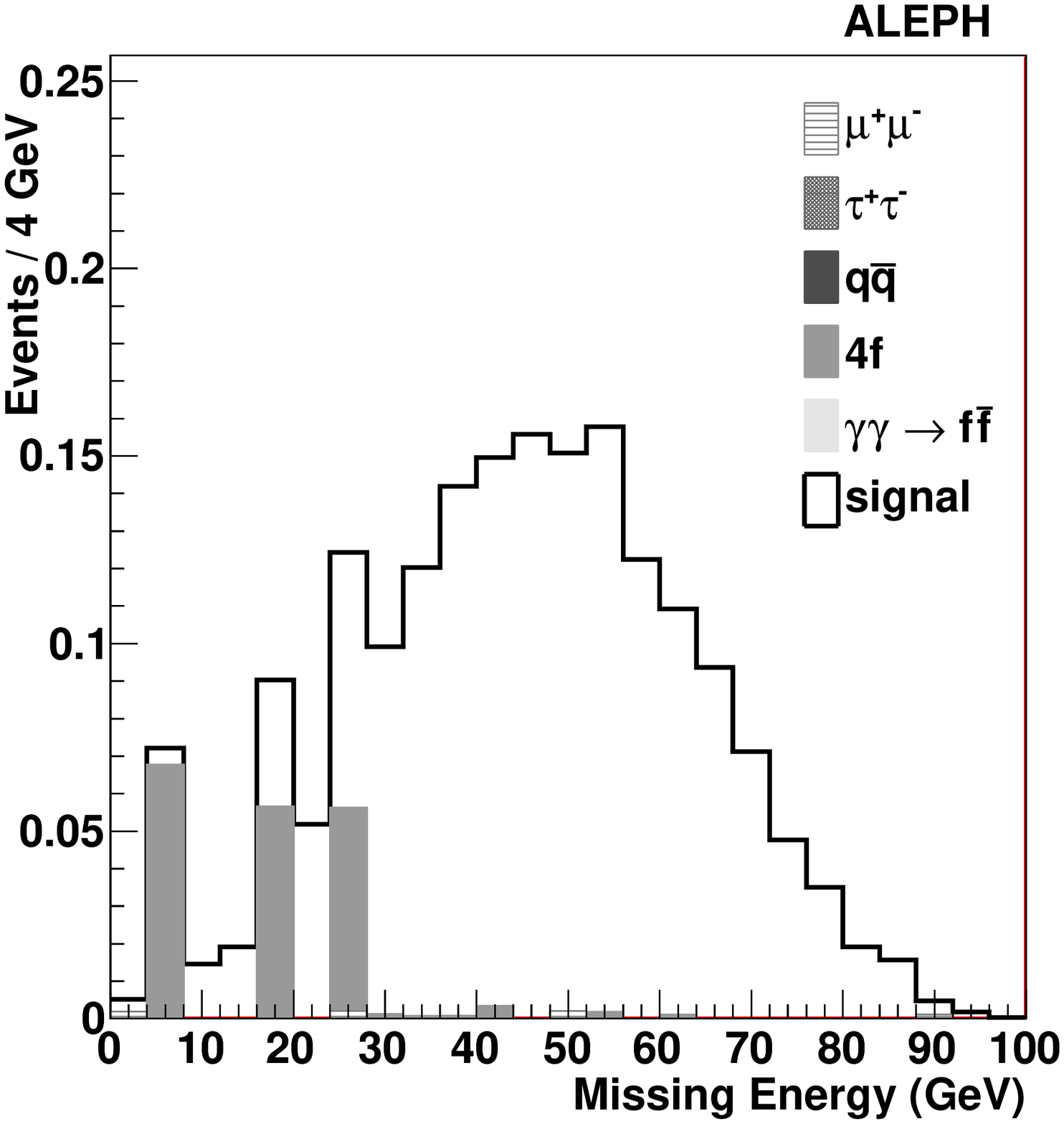}}
\end{center}
\caption{Distributions for the $\PZz\ra \mu^+\mu^-$ channel after the loose selection for (a) the reconstructed $\PZz$ invariant mass  and (b) missing energy, where signal corresponds to $\mh = 100~\GEVcc$, $\ma = 4~\GEVcc$ with $\xi^2=1$  (see text).  The same distributions are shown in (c) and (d) after the final selection, excluding any requirements on the variable shown.}
\label{fig:MuonPlots}
\end{figure}

The final selection consisted of the following requirements and maintained an acceptable signal efficiency while rejecting most backgrounds.  A mass window for the candidate $\PZz$ between $80$--$102 ~\GEVcc$ was effective at removing two-fermion backgrounds.  Due to the neutrinos from tau decays the signal was separated from fully hadronic final states by requiring $\slashed{E}> 20~\GEV$, where $\slashed{E}$ is the missing energy in the event.   The expected jet configuration of the signal was enforced by requiring  $\cos\theta_{\rm j_1j_2} < 0$, where $\theta_{\rm j_1j_2}$ is the angle between the two jets.  Finally, the remaining backgrounds were suppressed by requiring $\jtrk_{1,2} = 2$ or 4, the dominant track multiplicities expected in the signal.  Figures~\ref{fig:ElectronPlots} and ~\ref{fig:MuonPlots} show the distribution of the reconstructed $\PZz$ mass and missing energy for the $\PZz\ra \eplus\eminus$ and $\PZz\ra \mu^+\mu^-$ channels, respectively.  The numbers of events passing loose and final selection in data and simulated background are shown  in Table~\ref{tbl:CutFlow}.



\subsection{$\PZz\ra \nu\bar{\nu}$}

All objects found in the event were clustered into jets as described above. The loose selection consisted of the following requirements.  Missing energy greater than $30~\GEV$ and missing mass, $\slashed{m}$, greater than $20~\GEVcc$ were used to reject dijet and other two-fermion backgrounds.  In order to further reject the $\gamma\gamma$ background, events were required to have $E_{\rm vis} > 0.05 \, E_{\rm CM}$ and $|\cos\theta_{\rm me}|<0.9$, where $E_{\rm vis}$ is the visible energy and $\theta_{\rm me}$ is the angle between the missing momentum vector and the beam axis. 
 Events were required to have two well-contained jets with $|\cos\theta_{\rm j}| < 0.85$, dijet invariant mass $m_{\rm j_1 j_2} > 10~\GEVcc$, dijet angular separation $\cos\theta_{\rm j_1j_2} <0$, and the highest energy jet was required to have $E_{\rm j_1} > 25~\GEV$ and $\jtrk_1 = 2$ or  $4$.  
 
\begin{figure}[htb]
\begin{center}
\subfigure[]{\includegraphics[scale=.41]{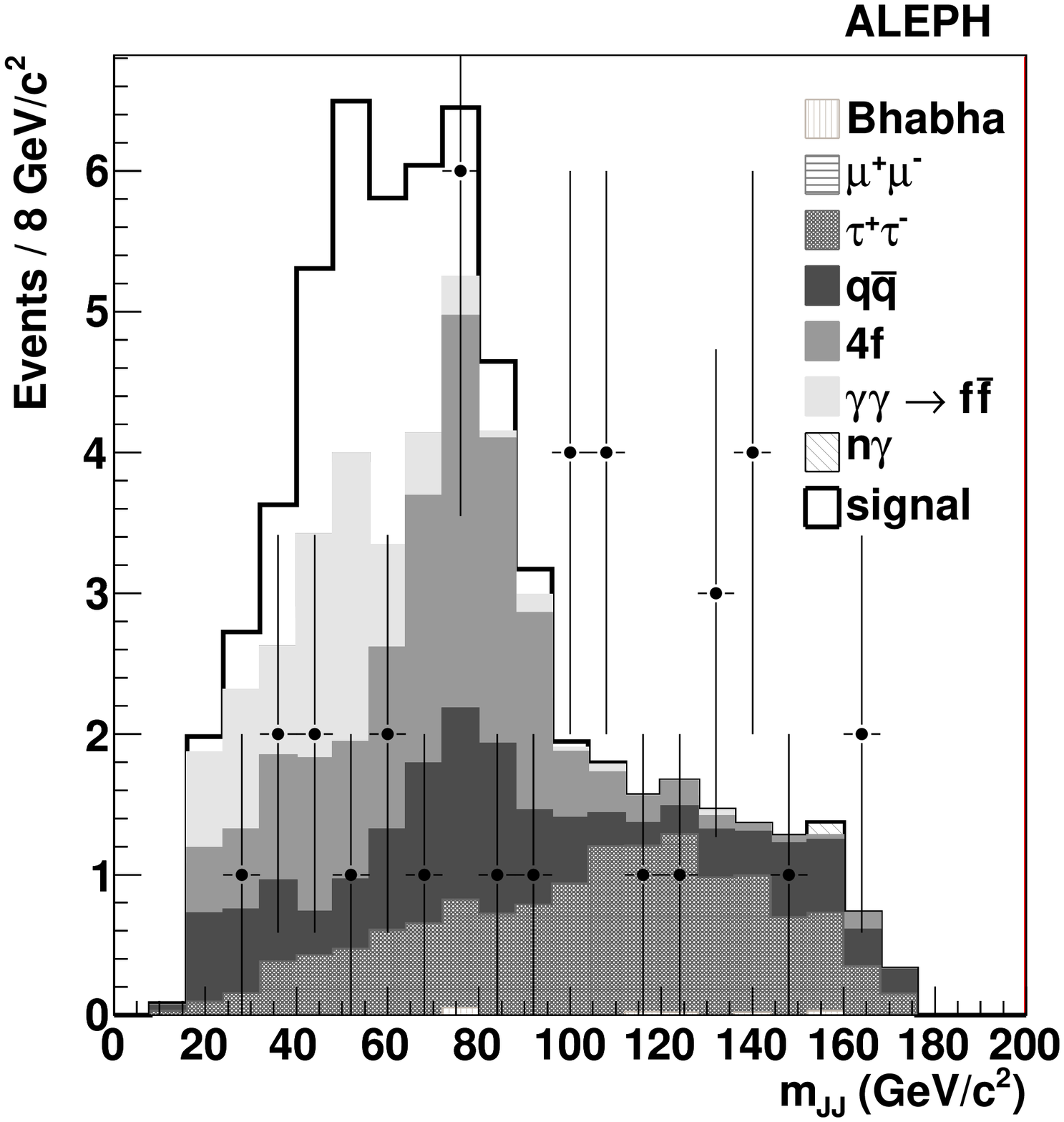}}%
\subfigure[]{\includegraphics[scale=.41]{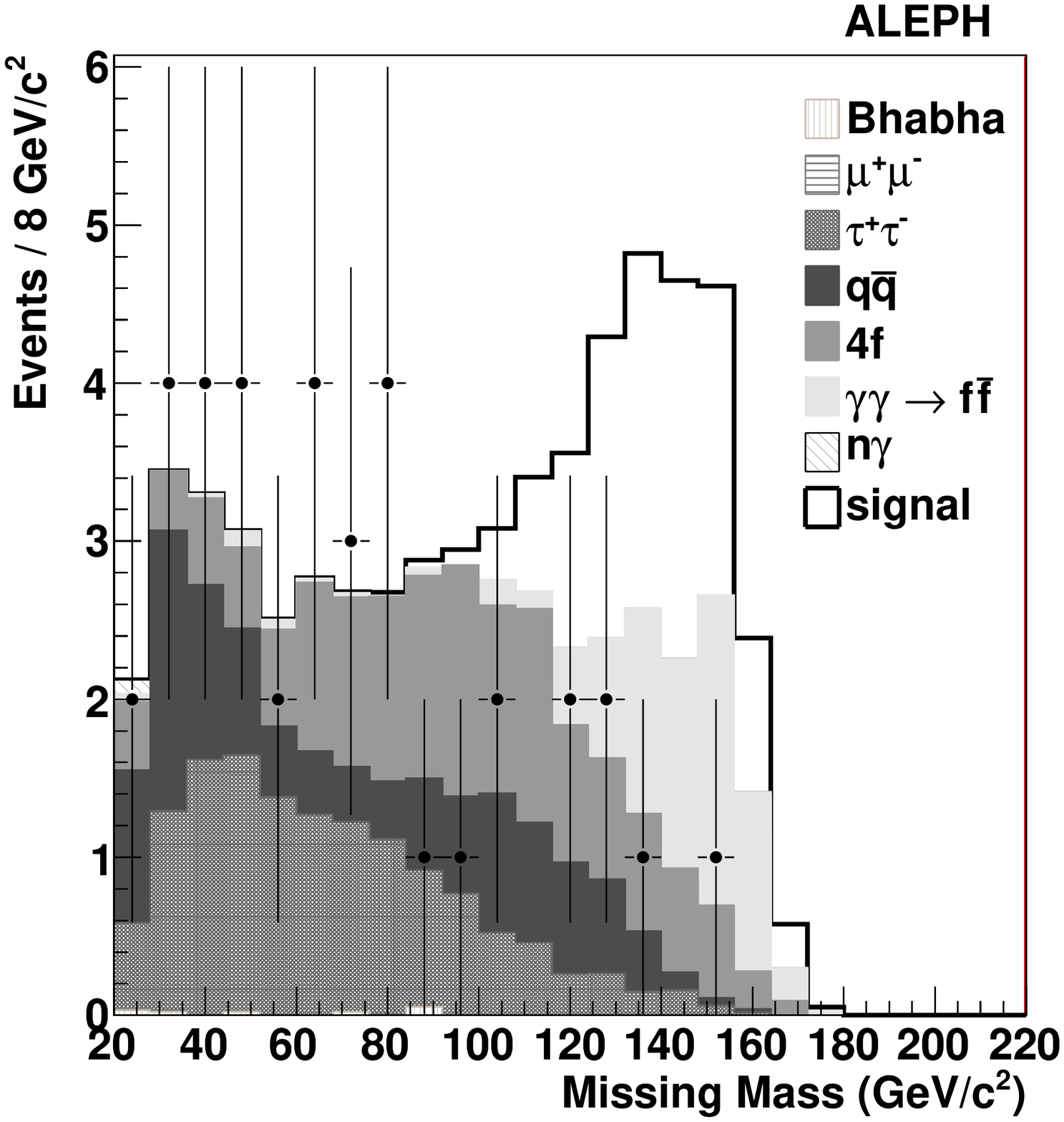}}\\
\subfigure[]{\includegraphics[scale=.41]{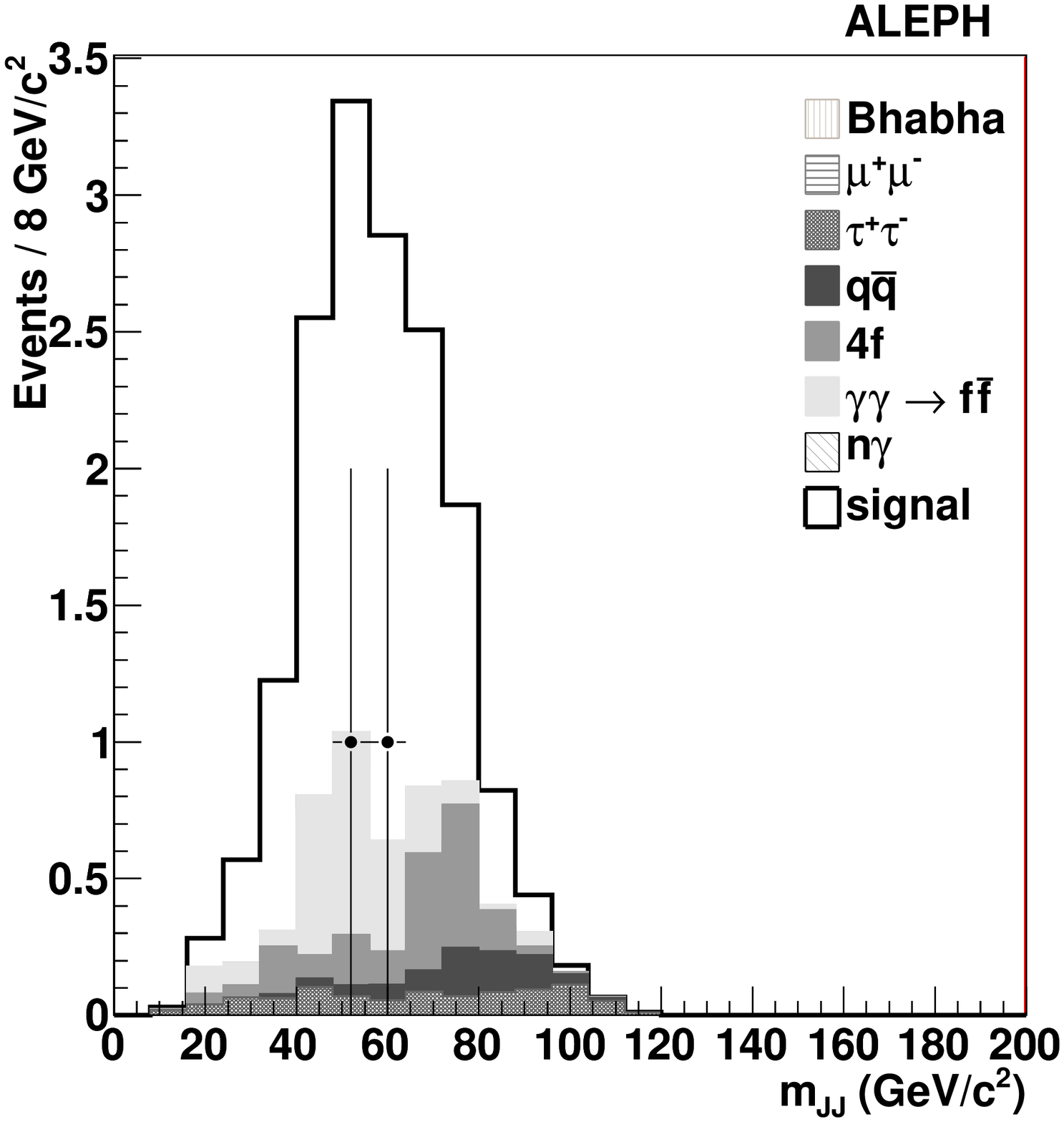}}%
\subfigure[]{\includegraphics[scale=.41]{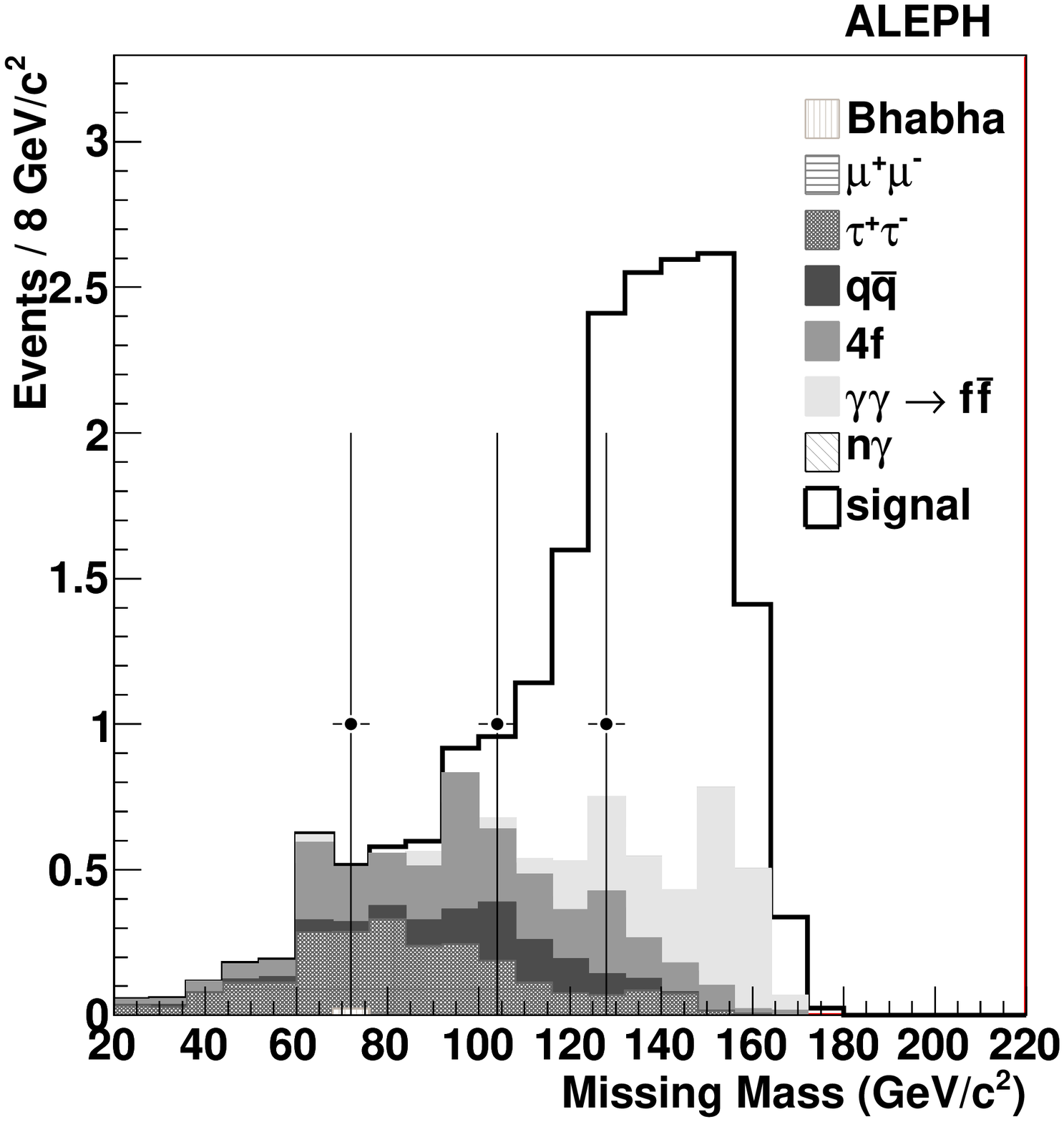}}
\end{center}
\caption{Distributions for the $\PZz\ra \nu\bar{\nu}$ channel after the loose selection and requirement of $1<\jtrk_2 < 7$ for  (a) dijet invariant mass and (b) missing mass, where signal corresponds to $\mh = 100~\GEVcc$, $\ma = 4~\GEVcc$ with $\xi^2=1$ (see text).  The same distributions are shown in (c) and (d) after the final selection, excluding any requirements on the variable shown.}
\label{fig:NeutrinoPlots}
\end{figure}

 The final selection consisted of the following requirements.  First, the requirement $E_{\rm j_1}+ E_{\rm j_2}+\slashed{E} > E_{\rm CM} - 5~\GEV$ was used to reject events with energy deposits in the forward regions of the detector. Consistency with $\PZz\ra \nu\bar{\nu}$ was ensured by requiring $\slashed{E}>60~\GEV$ and  $\slashed{m}>90~\GEVcc$.  The distribution of aplanarity for the signal is strongly peaked near 0, while the remaining backgrounds have a longer tail.  The tail of the aplanarity distribution for the signal extends further for larger $\ma$ and smaller $\mh$; larger $\ma$ leads to broader jets and lighter Higgs bosons can be  produced with more momentum reducing the opening angle between the jets in the laboratory frame.  Thus the requirement aplanarity$< 0.05$ was chosen to maintain an acceptable signal efficiency for $\mh=86~\GEVcc$ and $\ma=10~\GEVcc$.  Finally, the second jet was also required to have $\jtrk_{2} = 2$ or 4. Figure~\ref{fig:NeutrinoPlots} shows the distribution of dijet invariant mass and missing mass for the $\PZz\ra \nu\bar{\nu}$ channel.  The numbers of events passing loose and final selection in data and simulated background are shown  in Table~\ref{tbl:CutFlow}.

\begin{table}[htb]
\begin{center}
\caption{Number of events passing loose and final selections in each channel, in data, simulated background, and simulated signal ($\mh = 100~\GEVcc$, $\ma = 4~\GEVcc$).  The numbers of events passing the final selection are categorised  by track multiplicity. }
\label{tbl:CutFlow}
\vspace{3pt}
\scalebox{1}{
\begin{tabular}{|c||c|c|c|c|c|c|c|c|}
\hline
Channel & Selection & data & total & \multicolumn{4}{|c|}{background category} & signal \\
              &  $(\jtrk_1, \jtrk_2)$     &          & background & 2f & 4f & $\gamma\gamma$ & n$\gamma$ &  \\
\hline
\multirow{4}{*}{$\PZz\ra \eplus\eminus$} & Loose & 299 & 332 & 183 & 137 & 12.31 & 0.65 & 2.27 \\
& (2,2) & 0 & 0.034 & 0.034 & 0.000 & 0.000 & 0.000 & 0.689 \\
& (2,4)+(4,2) & 0 & 0.055 & 0.014 & 0.005 & 0.037 & 0.000 & 0.610 \\
&(4,4) & 0 & 0.031 & 0.019 & 0.013 & 0.000 & 0.000 & 0.126 \\
\hline
\hline
\multirow{4}{*}{$\PZz\ra \mu^+\mu^-$}  & Loose & 83 & 74.50 & 12.79 & 60.64 & 1.07 & 0.00 & 2.37 \\
&(2,2) & 0 & 0.058 & 0.005 & 0.053 & 0.000 & 0.000& 0.800 \\
&(2,4)+(4,2) & 0 & 0.005 & 0.000 & 0.005 & 0.000 & 0.000& 0.676 \\
&(4,4) & 0 & 0.006 & 0.000 & 0.006 & 0.000 & 0.000& 0.127 \\
\hline
\hline
\multirow{4}{*}{$\PZz\ra \nu\bar{\nu}$}  & Loose  & 206 & 200 & 135 & 47.97 & 13.50 & 3.74 & 12.63 \\
& (2,2) & 0 & 1.312 & 0.663 & 0.408 & 0.240 & 0.000 & 5.097 \\
& (2,4)+(4,2) & 0 & 1.948 & 0.528 & 0.575 & 0.845 & 0.000 & 4.741 \\
& (4,4) & 2 & 2.569 & 0.461 & 0.820 & 1.288 & 0.000 & 1.089 \\
\hline
\end{tabular}
}
\end{center}
\end{table}




\subsection{Signal Efficiency}

\begin{figure}[htb]
\begin{center}
\includegraphics[scale=.4]{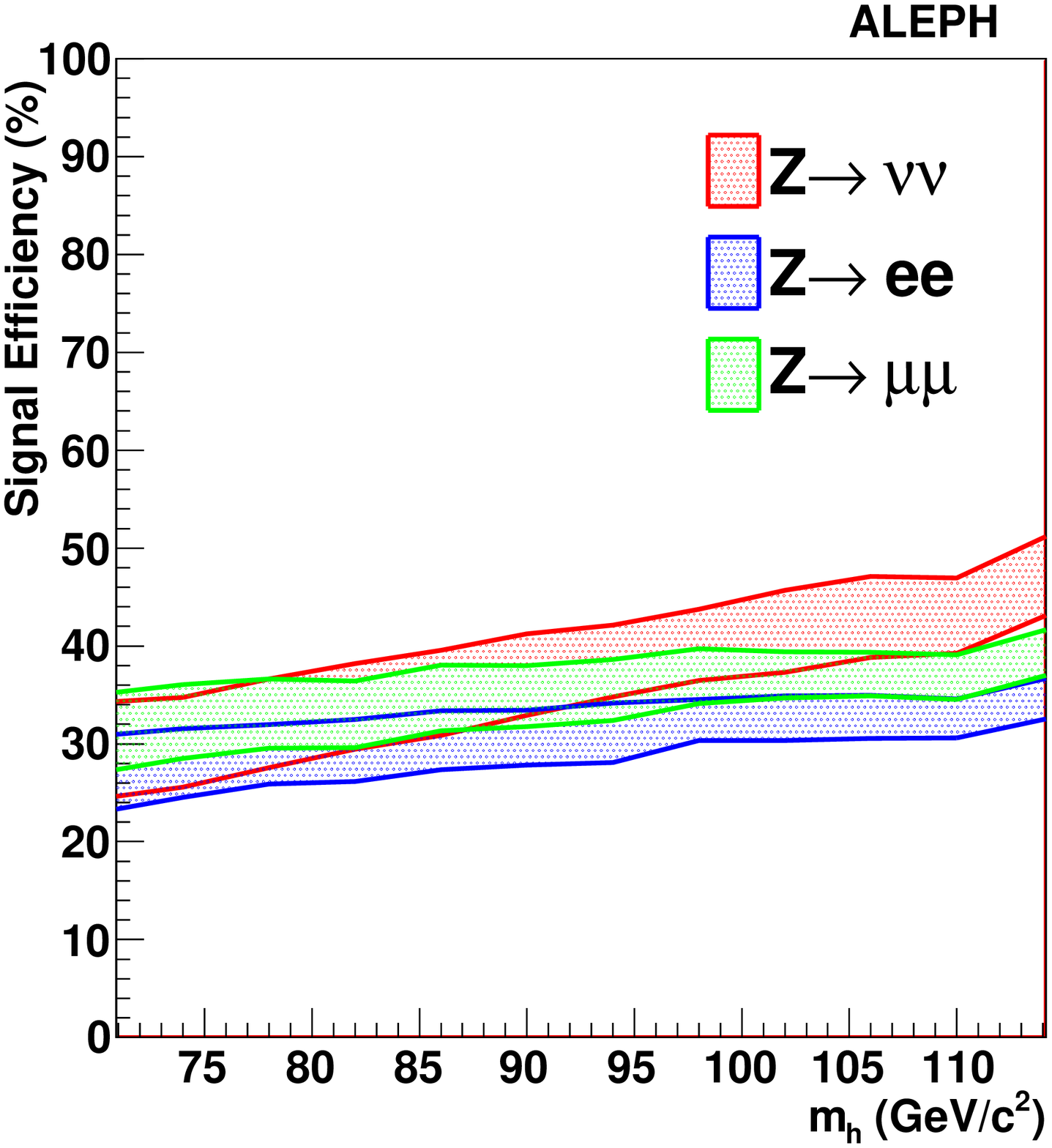}
\end{center}
\caption{Signal efficiency as a function of the Higgs boson mass for the three channels considered in this work, $\PZz\ra \eplus\eminus,~\mu^+\mu^-,$ and $\nu\bar{\nu}$.  The upper (lower) portion of the efficiency band corresponds to $\ma = 4 ~(10)~\GEVcc$.}
\label{fig:SigEfficiency}
\end{figure}

The $\mathrm{h}\ra 2\A \ra 4\tau$ signal efficiency is shown in Fig.~\ref{fig:SigEfficiency} as a function of the Higgs boson mass with $\ma = 4$--$10~\GEVcc$ for the three $\PZz$ decay channels considered. The decrease in efficiency at higher $\ma$ values, seen in Fig.~\ref{fig:SigEfficiency}, is due to two effects. First, the invariant mass of the jet becomes larger and the fraction of three-jet events increases. Second, and more importantly, the events become more aplanar.  


\section{Systematic Uncertainties}
\label{sec:systematics}

Uncertainties and inaccuracies in the Monte Carlo simulation lead to systematic effects in the analysis.  The impact of uncertainties in jet energy and direction, missing energy, and lepton identification and isolation were estimated in Ref.~\cite{Barate:2000na} for similar final states.  Compared to those, the present analyses do not use neural networks for event selection, do not use b tagging or tau tagging, and the simulated background samples are substantially larger. The systematic uncertainties related to the simulation of jet energies and directions were evaluated from the sample of hadronic events collected at the Z peak in 1998.  Based on that sample, it was found that additional smearing of the Monte Carlo simulation was not necessary in the barrel region of the detector.   

For the $\PZz \ra \ell^+ \ell^- $ channels, the total relative systematic uncertainties from lepton identification and isolation were found to be 0.6\%, 2.6\% and 7.5\% for the signal, $\PZz \PZz$, and $\PZz \mathrm{ee}$ backgrounds, respectively.    The systematic uncertainties for
$\mathrm{WW}, ~\mathrm{We}\nu$, $\qqbar$, and other backgrounds were all smaller than 30\%. Based on these estimates and the background composition, a 10\% uncertainty is estimated for the background in the $\PZz \ra \ell^+\ell^-$ channels. 

The cuts used for the $\PZz\ra\nu\bar{\nu}$ final state are sensitive to beam related backgrounds. The energy distribution of this background was measured with events recorded at random beam crossings. Additional energy depositions at angles below 12$^\circ$ were added randomly to all simulated events according to this energy distribution. The relative uncertainty in the total selection efficiency for the analyses presented in Ref.~\cite{Barate:2000na} was 5\% for the signal and 10\% for $\PZz \PZz$, and it is between 30\% and 100\% for the other background processes.  Based on these estimates and the background composition, the uncertainty for the background in the $\PZz \ra \nu\bar{\nu}$ channel is estimated to be 30\%.  

The agreement between the background estimate and the observed number of events in data with the loose selection is within the systematic and statistical uncertainty for all three channels. Given the low numbers of selected events, the final measurements are statistically limited.

\section{Results}
\label{sec:results}

No excess of events above the background was observed. Limits on the cross section times branching ratio with respect to the SM Higgsstrahlung production cross section, $\xi^2 = \frac{\sigma(\eplus \eminus\ra \PZz \mathrm{h})}{\sigma_{\rm SM}(\eplus \eminus\ra \PZz\mathrm{h})}\times B(\mathrm{h}\ra \A\A)\times B(\A\ra \tau^+\tau^-)^2$ were determined as follows.  A joint probability density was constructed to describe the number of events in each of three jet-multiplicity pairings (indexed by $m$) for each of the three final states (indexed by $f$).  The three jet-multiplicity pairings correspond to the different permutations of one-prong and three-prong tau decays in each of the jets, ignoring those with six tracks in an individual jet, leaving the three permutations $(\jtrk_1, \jtrk_2) \in  \{ (2,2), (2,4) \textrm{ or } (4,2), (4,4)\} \equiv  \mathcal{M}$.  The event count $N_{m,f}$ in each of these nine categories was modeled with a Poisson distribution about the sum of the uncertain background $b_{m,f}$ and the expected signal $s_{m,f}$ scaled by $\xi^2$.  A normal distribution was used to model the relationship between the uncertain background, the Monte Carlo-based background estimate $b_{m,f}^{\rm MC}$, and its systematic uncertainty $\Delta_f$.  This procedure leads to the following joint probability density for the event counts:
\begin{equation}
P( N_{m,f} | \xi^2, b_{m,f}) = \prod_{m \in \mathcal{M}} \, \prod_{f \in \{ee, \mu\mu, \nu\nu\}} {\rm Pois}(N_{m,f} | \xi^2 s_{m,f} + b_{m,f}) \cdot N(b_{m,f}^{\rm MC}| b_{m,f}, \Delta_{f}).
\end{equation}

\begin{figure}[h]
\begin{center}
\subfigure[]{\includegraphics[scale=.4]{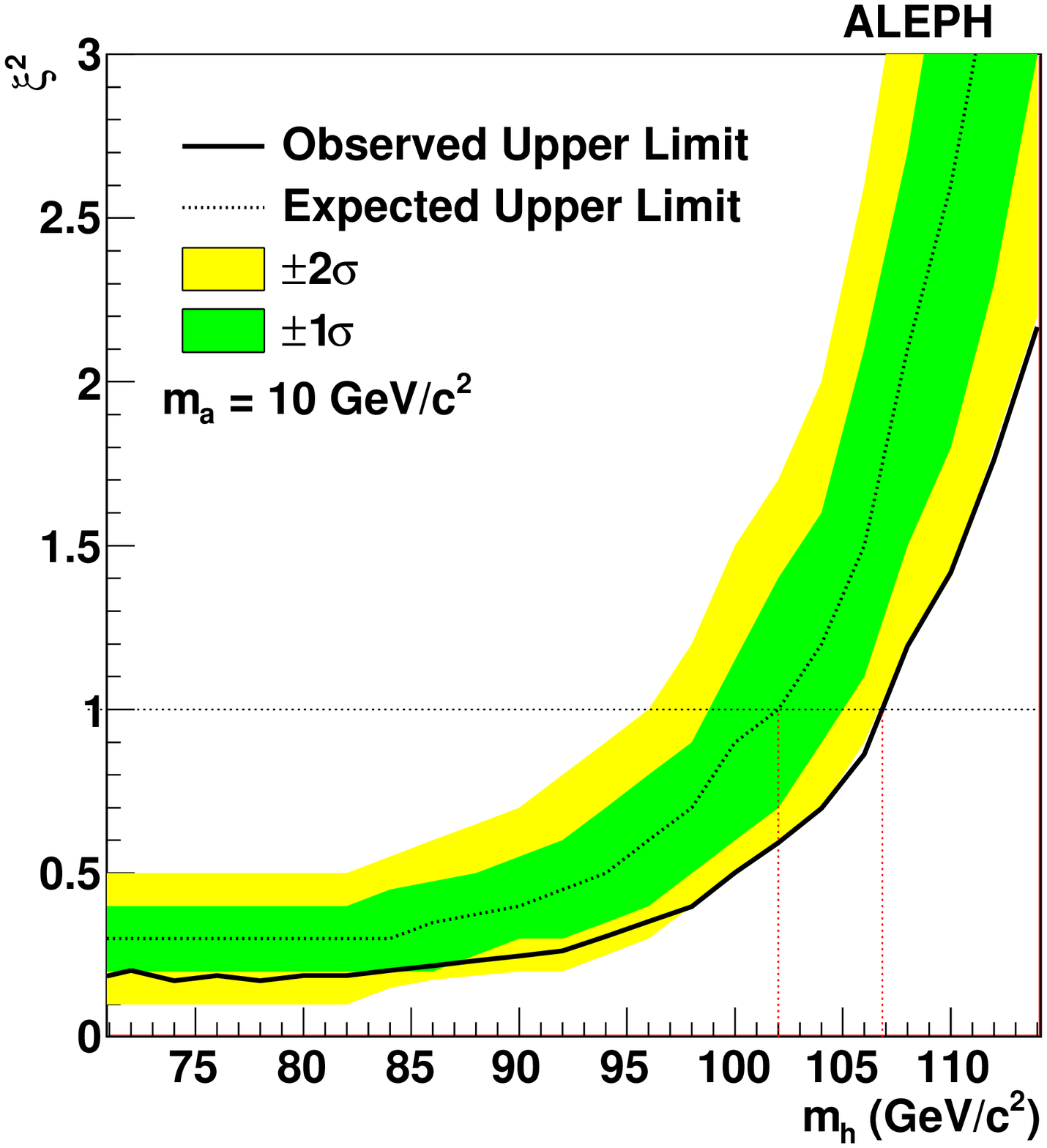}}%
\subfigure[]{\includegraphics[scale=.4]{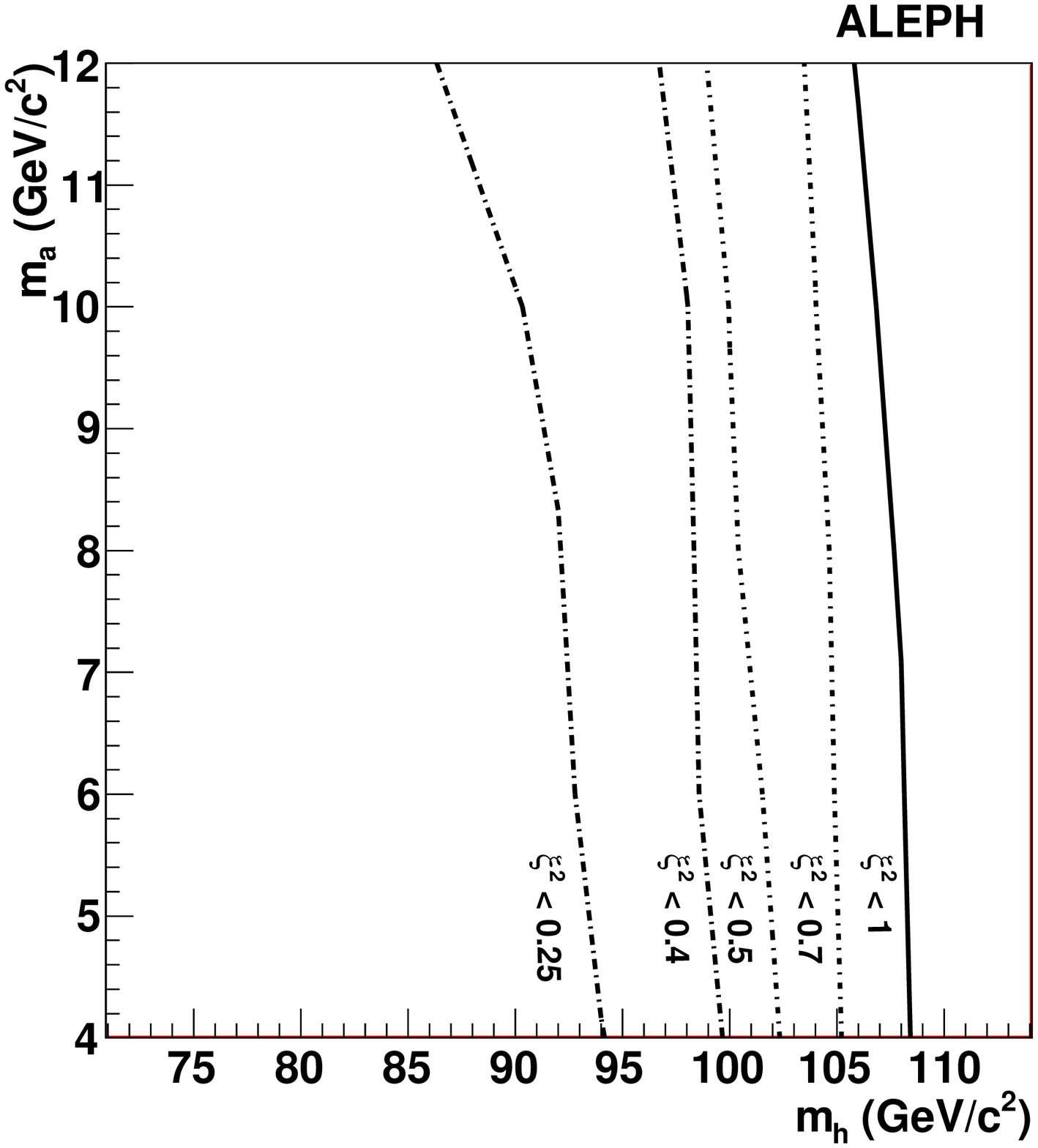}}
\end{center}
\caption{(a) Observed and expected 95\% confidence level limit on $\xi^2$ as a function of the Higgs boson mass for $\ma=10~\GEVcc$.  (b) Contours of observed 95\% confidence level limit on $\xi^2$ in the $(\mh,\ma)$  plane.}
\label{fig:limit_band}
\end{figure}

Confidence intervals were constructed by using  a generalized version of the Feldman-Cousins technique~\cite{feldman-1998-57}, which incorporates systematic uncertainties in a frequentist way~\cite{chuang, Cranmer:2005hi}.   Figure~\ref{fig:limit_band}a shows the 95\% confidence level upper-limit on $\xi^2$ as a function of $\mh$ for $\ma = 10~\GEVcc$. Figure \ref{fig:limit_band}b shows 95\% confidence level contours of $\xi^2$ in  the $(\mh,\ma)$ plane.  Because the selection has no $\mh$ or $\ma$ dependence, the resulting upper limits are fully correlated.  The observed number of events is consistent with a downward fluctuation of the background, which leads to stronger than expected limits on $\xi^2$.


\section{Conclusions}
\label{sec:conclusions}

A search for a Higgs boson produced via Higgsstrahlung at LEP2 energies has been performed, where $\mathrm{h}\ra 2\A\ra 4\tau$ and $\PZz\ra \eplus\eminus,\mu^+\mu^-,$ $\nu\bar{\nu}$. No evidence for an excess of events above background was observed, and a limit on the combined production cross section times branching ratio, $\xi^2=\frac{\sigma(\eplus \eminus\ra \PZz \mathrm{h})}{\sigma_{\rm SM}(\eplus \eminus\ra \PZz \mathrm{h})}\times B(\mathrm{h}\ra \A\A)\times B(\A\ra \tau^+\tau^-)^2$ is presented. For $\mh< 107~\GEVcc$ and  $4<\ma<10~\GEVcc$, $\xi^2>1$ is excluded at the 95\% confidence level.
This analysis covers a region of parameter space previously left unexplored, and further constrains models with non-standard Higgs decays, such as the NMSSM.  

\section*{Acknowledgments}
We wish to thank Neal Weiner and Riccardo Barbieri for providing encouragement, motivation, and advice throughout this work. It is a pleasure to congratulate our colleagues from the CERN accelerator divisions
for the successful operation of LEP throughout the LEP2 years.
We are indebted to the engineers and technicians in all our institutions for
their contributions to the excellent performance of ALEPH.
Those of us from non-member countries thank CERN for its hospitality.

\bibliography{FourTaus}
\bibliographystyle{JHEP}
\end{document}

%% file: authb.tex
\pagestyle{empty}
\newpage
\small
%
%
\newlength{\saveparskip}
\newlength{\savetextheight}
\newlength{\savetopmargin}
\newlength{\savetextwidth}
\newlength{\saveoddsidemargin}
\newlength{\savetopsep}
\setlength{\saveparskip}{\parskip}
\setlength{\savetextheight}{\textheight}
\setlength{\savetopmargin}{\topmargin}
\setlength{\savetextwidth}{\textwidth}
\setlength{\saveoddsidemargin}{\oddsidemargin}
\setlength{\savetopsep}{\topsep}
%
%
\setlength{\parskip}{0.0cm}
\setlength{\textheight}{25.0cm}
\setlength{\topmargin}{-1.5cm}
\setlength{\textwidth}{16 cm}
\setlength{\oddsidemargin}{-0.0cm}
\setlength{\topsep}{1mm}
\pretolerance=10000
\centerline{\large\bf The ALEPH Collaboration}
\footnotesize
\vspace{0.5cm}
{\raggedbottom
\begin{sloppypar}
\samepage\noindent
S.~Schael
\nopagebreak
\begin{center}
\parbox{15.5cm}{\sl\samepage
Physikalisches Institut der RWTH-Aachen, D-52056 Aachen, Germany}
\end{center}\end{sloppypar}
\vspace{2mm}
\begin{sloppypar}
\noindent
R.~Barate,
R.~Bruneli\`ere,
I.~De~Bonis,
D.~Decamp,
C.~Goy,
S.~J\'ez\'equel,
J.-P.~Lees,
F.~Martin,
E.~Merle,
\mbox{M.-N.~Minard},
B.~Pietrzyk,
B.~Trocm\'e
\nopagebreak
\begin{center}
\parbox{15.5cm}{\sl\samepage
Laboratoire de Physique des Particules (LAPP), IN$^{2}$P$^{3}$-CNRS,
F-74019 Annecy-le-Vieux Cedex, France}
\end{center}\end{sloppypar}
\vspace{2mm}
\begin{sloppypar}
\noindent
S.~Bravo,
M.P.~Casado,
M.~Chmeissani,
J.M.~Crespo,
E.~Fernandez,
M.~Fernandez-Bosman,
Ll.~Garrido,$^{15}$
M.~Martinez,
A.~Pacheco,
H.~Ruiz
\nopagebreak
\begin{center}
\parbox{15.5cm}{\sl\samepage
Institut de F\'{i}sica d'Altes Energies, Universitat Aut\`{o}noma
de Barcelona, E-08193 Bellaterra (Barcelona), Spain$^{7}$}
\end{center}\end{sloppypar}
\vspace{2mm}
\begin{sloppypar}
\noindent
A.~Colaleo,
D.~Creanza,
N.~De~Filippis,
M.~de~Palma,
G.~Iaselli,
G.~Maggi,
M.~Maggi,
S.~Nuzzo,
A.~Ranieri,
G.~Raso,$^{24}$
F.~Ruggieri,
G.~Selvaggi,
L.~Silvestris,
P.~Tempesta,
A.~Tricomi,$^{3}$
G.~Zito
\nopagebreak
\begin{center}
\parbox{15.5cm}{\sl\samepage
Dipartimento di Fisica, INFN Sezione di Bari, I-70126 Bari, Italy}
\end{center}\end{sloppypar}
\vspace{2mm}
\begin{sloppypar}
\noindent
X.~Huang,
J.~Lin,
Q. Ouyang,
T.~Wang,
Y.~Xie,
R.~Xu,
S.~Xue,
J.~Zhang,
L.~Zhang,
W.~Zhao
\nopagebreak
\begin{center}
\parbox{15.5cm}{\sl\samepage
Institute of High Energy Physics, Academia Sinica, Beijing, The People's
Republic of China$^{8}$}
\end{center}\end{sloppypar}
\vspace{2mm}
\begin{sloppypar}
\noindent
D.~Abbaneo,
T.~Barklow,$^{26}$
O.~Buchm\"uller,$^{26}$
M.~Cattaneo,
B.~Clerbaux,$^{23}$
H.~Drevermann,
R.W.~Forty,
M.~Frank,
F.~Gianotti,
J.B.~Hansen,
J.~Harvey,
D.E.~Hutchcroft,$^{30}$,
P.~Janot,
B.~Jost,
M.~Kado,$^{2}$
P.~Mato,
A.~Moutoussi,
F.~Ranjard,
L.~Rolandi,
D.~Schlatter,
F.~Teubert,
A.~Valassi,
I.~Videau
\nopagebreak
\begin{center}
\parbox{15.5cm}{\sl\samepage
European Laboratory for Particle Physics (CERN), CH-1211 Geneva 23,
Switzerland}
\end{center}\end{sloppypar}
\vspace{2mm}
\begin{sloppypar}
\noindent
F.~Badaud,
S.~Dessagne,
A.~Falvard,$^{20}$
D.~Fayolle,
P.~Gay,
J.~Jousset,
B.~Michel,
S.~Monteil,
D.~Pallin,
J.M.~Pascolo,
P.~Perret
\nopagebreak
\begin{center}
\parbox{15.5cm}{\sl\samepage
Laboratoire de Physique Corpusculaire, Universit\'e Blaise Pascal,
IN$^{2}$P$^{3}$-CNRS, Clermont-Ferrand, F-63177 Aubi\`{e}re, France}
\end{center}\end{sloppypar}
\vspace{2mm}
\begin{sloppypar}
\noindent
J.D.~Hansen,
J.R.~Hansen,
P.H.~Hansen,
A.C.~Kraan,
B.S.~Nilsson
\nopagebreak
\begin{center}
\parbox{15.5cm}{\sl\samepage
Niels Bohr Institute, 2100 Copenhagen, DK-Denmark$^{9}$}
\end{center}\end{sloppypar}
\vspace{2mm}
\begin{sloppypar}
\noindent
A.~Kyriakis,
C.~Markou,
E.~Simopoulou,
A.~Vayaki,
K.~Zachariadou
\nopagebreak
\begin{center}
\parbox{15.5cm}{\sl\samepage
Nuclear Research Center Demokritos (NRCD), GR-15310 Attiki, Greece}
\end{center}\end{sloppypar}
\vspace{2mm}
\begin{sloppypar}
\noindent
A.~Blondel,$^{12}$
\mbox{J.-C.~Brient},
F.~Machefert,
A.~Roug\'{e},
H.~Videau
\nopagebreak
\begin{center}
\parbox{15.5cm}{\sl\samepage
Laoratoire Leprince-Ringuet, Ecole
Polytechnique, IN$^{2}$P$^{3}$-CNRS, \mbox{F-91128} Palaiseau Cedex, France}
\end{center}\end{sloppypar}
\vspace{2mm}
\begin{sloppypar}
\noindent
V.~Ciulli,
E.~Focardi,
G.~Parrini
\nopagebreak
\begin{center}
\parbox{15.5cm}{\sl\samepage
Dipartimento di Fisica, Universit\`a di Firenze, INFN Sezione di Firenze,
I-50125 Firenze, Italy}
\end{center}\end{sloppypar}
\vspace{2mm}
\begin{sloppypar}
\noindent
A.~Antonelli,
M.~Antonelli,
G.~Bencivenni,
F.~Bossi,
G.~Capon,
F.~Cerutti,
V.~Chiarella,
P.~Laurelli,
G.~Mannocchi,$^{5}$
G.P.~Murtas,
L.~Passalacqua
\nopagebreak
\begin{center}
\parbox{15.5cm}{\sl\samepage
Laboratori Nazionali dell'INFN (LNF-INFN), I-00044 Frascati, Italy}
\end{center}\end{sloppypar}
\vspace{2mm}
\begin{sloppypar}
\noindent
J.~Kennedy,
J.G.~Lynch,
P.~Negus,
V.~O'Shea,
A.S.~Thompson
\nopagebreak
\begin{center}
\parbox{15.5cm}{\sl\samepage
Department of Physics and Astronomy, University of Glasgow, Glasgow G12
8QQ,United Kingdom$^{10}$}
\end{center}\end{sloppypar}
\vspace{2mm}
\begin{sloppypar}
\noindent
S.~Wasserbaech
\nopagebreak
\begin{center}
\parbox{15.5cm}{\sl\samepage
Utah Valley State College, Orem, UT 84058, U.S.A.}
\end{center}\end{sloppypar}
\vspace{2mm}
\begin{sloppypar}
\noindent
R.~Cavanaugh,$^{4}$
S.~Dhamotharan,$^{21}$
C.~Geweniger,
P.~Hanke,
V.~Hepp,
E.E.~Kluge,
A.~Putzer,
H.~Stenzel,
K.~Tittel,
M.~Wunsch$^{19}$
\nopagebreak
\begin{center}
\parbox{15.5cm}{\sl\samepage
Kirchhoff-Institut f\"ur Physik, Universit\"at Heidelberg, D-69120
Heidelberg, Germany$^{16}$}
\end{center}\end{sloppypar}
\vspace{2mm}
\begin{sloppypar}
\noindent
R.~Beuselinck,
W.~Cameron,
G.~Davies,
P.J.~Dornan,
M.~Girone,$^{1}$
N.~Marinelli,
J.~Nowell,
S.A.~Rutherford,
J.K.~Sedgbeer,
J.C.~Thompson,$^{14}$
R.~White
\nopagebreak
\begin{center}
\parbox{15.5cm}{\sl\samepage
Department of Physics, Imperial College, London SW7 2BZ,
United Kingdom$^{10}$}
\end{center}\end{sloppypar}
\vspace{2mm}
\begin{sloppypar}
\noindent
V.M.~Ghete,
P.~Girtler,
E.~Kneringer,
D.~Kuhn,
G.~Rudolph
\nopagebreak
\begin{center}
\parbox{15.5cm}{\sl\samepage
Institut f\"ur Experimentalphysik, Universit\"at Innsbruck, A-6020
Innsbruck, Austria$^{18}$}
\end{center}\end{sloppypar}
\vspace{2mm}
\begin{sloppypar}
\noindent
E.~Bouhova-Thacker,
C.K.~Bowdery,
D.P.~Clarke,
G.~Ellis,
A.J.~Finch,
F.~Foster,
G.~Hughes,
R.W.L.~Jones,
M.R.~Pearson,
N.A.~Robertson,
T.~Sloan,
M.~Smizanska
\nopagebreak
\begin{center}
\parbox{15.5cm}{\sl\samepage
Department of Physics, University of Lancaster, Lancaster LA1 4YB,
United Kingdom$^{10}$}
\end{center}\end{sloppypar}
\vspace{2mm}
\begin{sloppypar}
\noindent
O.~van~der~Aa,
C.~Delaere,$^{28}$
G.Leibenguth,$^{31}$
V.~Lemaitre$^{29}$
\nopagebreak
\begin{center}
\parbox{15.5cm}{\sl\samepage
Institut de Physique Nucl\'eaire, D\'epartement de Physique, Universit\'e Catholique de Louvain, 1348 Louvain-la-Neuve, Belgium}
\end{center}\end{sloppypar}
\vspace{2mm}
\begin{sloppypar}
\noindent
U.~Blumenschein,
F.~H\"olldorfer,
K.~Jakobs,
F.~Kayser,
A.-S.~M\"uller,
B.~Renk,
H.-G.~Sander,
S.~Schmeling,
H.~Wachsmuth,
C.~Zeitnitz,
T.~Ziegler
\nopagebreak
\begin{center}
\parbox{15.5cm}{\sl\samepage
Institut f\"ur Physik, Universit\"at Mainz, D-55099 Mainz, Germany$^{16}$}
\end{center}\end{sloppypar}
\vspace{2mm}
\begin{sloppypar}
\noindent
A.~Bonissent,
P.~Coyle,
C.~Curtil,
A.~Ealet,
D.~Fouchez,
P.~Payre,
A.~Tilquin
\nopagebreak
\begin{center}
\parbox{15.5cm}{\sl\samepage
Centre de Physique des Particules de Marseille, Univ M\'editerran\'ee,
IN$^{2}$P$^{3}$-CNRS, F-13288 Marseille, France}
\end{center}\end{sloppypar}
\vspace{2mm}
\begin{sloppypar}
\noindent
F.~Ragusa
\nopagebreak
\begin{center}
\parbox{15.5cm}{\sl\samepage
Dipartimento di Fisica, Universit\`a di Milano e INFN Sezione di
Milano, I-20133 Milano, Italy.}
\end{center}\end{sloppypar}
\vspace{2mm}
\begin{sloppypar}
\noindent
A.~David,
H.~Dietl,$^{32}$
G.~Ganis,$^{27}$
K.~H\"uttmann,
G.~L\"utjens,
W.~M\"anner$^{32}$,
\mbox{H.-G.~Moser},
R.~Settles,
M.~Villegas,
G.~Wolf
\nopagebreak
\begin{center}
\parbox{15.5cm}{\sl\samepage
Max-Planck-Institut f\"ur Physik, Werner-Heisenberg-Institut,
D-80805 M\"unchen, Germany\footnotemark[16]}
\end{center}\end{sloppypar}
\vspace{2mm}
\begin{sloppypar}
\noindent
J.~Beacham, 
K.~Cranmer\footnotemark[33],
I.~Yavin\footnotemark[34]
\nopagebreak
\begin{center}
\parbox{15.5cm}{\sl\samepage
Center for Cosmology and Particle Physics, New York University, 
New York, NY 10003, USA}
\end{center}\end{sloppypar}
\vspace{2mm}
\begin{sloppypar}
\noindent
J.~Boucrot,
O.~Callot,
M.~Davier,
L.~Duflot,
\mbox{J.-F.~Grivaz},
Ph.~Heusse,
A.~Jacholkowska,$^{6}$
L.~Serin,
\mbox{J.-J.~Veillet}
\nopagebreak
\begin{center}
\parbox{15.5cm}{\sl\samepage
Laboratoire de l'Acc\'el\'erateur Lin\'eaire, Universit\'e de Paris-Sud,
IN$^{2}$P$^{3}$-CNRS, F-91898 Orsay Cedex, France}
\end{center}\end{sloppypar}
\vspace{2mm}
\begin{sloppypar}
\noindent
P.~Azzurri, 
G.~Bagliesi,
T.~Boccali,
L.~Fo\`a,
A.~Giammanco,
A.~Giassi,
F.~Ligabue,
A.~Messineo,
F.~Palla,
G.~Sanguinetti,
A.~Sciab\`a,
G.~Sguazzoni,
P.~Spagnolo,
R.~Tenchini,
A.~Venturi,
P.G.~Verdini
\samepage
\begin{center}
\parbox{15.5cm}{\sl\samepage
Dipartimento di Fisica dell'Universit\`a, INFN Sezione di Pisa,
e Scuola Normale Superiore, I-56010 Pisa, Italy}
\end{center}\end{sloppypar}
\vspace{2mm}
\begin{sloppypar}
\noindent
O.~Awunor,
G.A.~Blair,
G.~Cowan,
A.~Garcia-Bellido,
M.G.~Green,
T.~Medcalf,$^{25}$
A.~Misiejuk,
J.A.~Strong,$^{25}$
P.~Teixeira-Dias
\nopagebreak
\begin{center}
\parbox{15.5cm}{\sl\samepage
Department of Physics, Royal Holloway \& Bedford New College,
University of London, Egham, Surrey TW20 OEX, United Kingdom$^{10}$}
\end{center}\end{sloppypar}
\vspace{2mm}
\begin{sloppypar}
\noindent
R.W.~Clifft,
T.R.~Edgecock,
P.R.~Norton,
I.R.~Tomalin,
J.J.~Ward
\nopagebreak
\begin{center}
\parbox{15.5cm}{\sl\samepage
Particle Physics Dept., Rutherford Appleton Laboratory,
Chilton, Didcot, Oxon OX11 OQX, United Kingdom$^{10}$}
\end{center}\end{sloppypar}
\vspace{2mm}
\begin{sloppypar}
\noindent
\mbox{B.~Bloch-Devaux},
D.~Boumediene,
P.~Colas,
B.~Fabbro,
E.~Lan\c{c}on,
\mbox{M.-C.~Lemaire},
E.~Locci,
P.~Perez,
J.~Rander,
B.~Tuchming,
B.~Vallage
\nopagebreak
\begin{center}
\parbox{15.5cm}{\sl\samepage
CEA, DAPNIA/Service de Physique des Particules,
CE-Saclay, F-91191 Gif-sur-Yvette Cedex, France$^{17}$}
\end{center}\end{sloppypar}
\vspace{2mm}
\begin{sloppypar}
\noindent
A.M.~Litke,
G.~Taylor
\nopagebreak
\begin{center}
\parbox{15.5cm}{\sl\samepage
Institute for Particle Physics, University of California at
Santa Cruz, Santa Cruz, CA 95064, USA$^{22}$}
\end{center}\end{sloppypar}
\vspace{2mm}
\begin{sloppypar}
\noindent
C.N.~Booth,
S.~Cartwright,
F.~Combley,$^{25}$
P.N.~Hodgson,
M.~Lehto,
L.F.~Thompson
\nopagebreak
\begin{center}
\parbox{15.5cm}{\sl\samepage
Department of Physics, University of Sheffield, Sheffield S3 7RH,
United Kingdom$^{10}$}
\end{center}\end{sloppypar}
\vspace{2mm}
\begin{sloppypar}
\noindent
A.~B\"ohrer,
S.~Brandt,
C.~Grupen,
J.~Hess,
A.~Ngac,
G.~Prange
\nopagebreak
\begin{center}
\parbox{15.5cm}{\sl\samepage
Fachbereich Physik, Universit\"at Siegen, D-57068 Siegen, Germany$^{16}$}
\end{center}\end{sloppypar}
\vspace{2mm}
\begin{sloppypar}
\noindent
C.~Borean,
G.~Giannini
\nopagebreak
\begin{center}
\parbox{15.5cm}{\sl\samepage
Dipartimento di Fisica, Universit\`a di Trieste e INFN Sezione di Trieste,
I-34127 Trieste, Italy}
\end{center}\end{sloppypar}
\vspace{2mm}
\begin{sloppypar}
\noindent
H.~He,
J.~Putz,
J.~Rothberg
\nopagebreak
\begin{center}
\parbox{15.5cm}{\sl\samepage
Experimental Elementary Particle Physics, University of Washington, Seattle,
WA 98195 U.S.A.}
\end{center}\end{sloppypar}
\vspace{2mm}
\begin{sloppypar}
\noindent
S.R.~Armstrong,
K.~Berkelman,
D.P.S.~Ferguson,
Y.~Gao,$^{13}$
S.~Gonz\'{a}lez,
O.J.~Hayes,
H.~Hu,
S.~Jin,
J.~Kile,
P.A.~McNamara III,
J.~Nielsen,
Y.B.~Pan,
\mbox{J.H.~von~Wimmersperg-Toeller}, 
W.~Wiedenmann,
J.~Wu,
Sau~Lan~Wu,
X.~Wu,
G.~Zobernig
\nopagebreak
\begin{center}
\parbox{15.5cm}{\sl\samepage
Department of Physics, University of Wisconsin, Madison, WI 53706,
USA$^{11}$}
\end{center}\end{sloppypar}
\vspace{2mm}
\begin{sloppypar}
\noindent
G.~Dissertori
\nopagebreak
\begin{center}
\parbox{15.5cm}{\sl\samepage
Institute for Particle Physics, ETH H\"onggerberg, 8093 Z\"urich,
Switzerland.}
\end{center}\end{sloppypar}
}
\footnotetext[1]{Also at CERN, 1211 Geneva 23, Switzerland.}
\footnotetext[2]{Now at Fermilab, PO Box 500, MS 352, Batavia, IL 60510, USA}
\footnotetext[3]{Also at Dipartimento di Fisica di Catania and INFN Sezione di
 Catania, 95129 Catania, Italy.}
\footnotetext[4]{Now at University of Florida, Department of Physics, Gainesville, Florida 32611-8440, USA}
\footnotetext[5]{Also IFSI sezione di Torino, INAF, Italy.}
\footnotetext[6]{Also at Groupe d'Astroparticules de Montpellier, Universit\'{e} de Montpellier II, Montpellier, France.}
\footnotetext[7]{Supported by CICYT, Spain.}
\footnotetext[8]{Supported by the National Science Foundation of China.}
\footnotetext[9]{Supported by the Danish Natural Science Research Council.}
\footnotetext[10]{Supported by the UK Particle Physics and Astronomy Research
Council.}
\footnotetext[11]{Supported by the US Department of Energy, grant
DE-FG0295-ER40896.}
\footnotetext[12]{Now at Departement de Physique Corpusculaire, Universit\'e de
Gen\`eve, 1211 Gen\`eve 4, Switzerland.}
\footnotetext[13]{Also at Department of Physics, Tsinghua University, Beijing, The People's Republic of China.}
\footnotetext[14]{Supported by the Leverhulme Trust.}
\footnotetext[15]{Permanent address: Universitat de Barcelona, 08208 Barcelona,
Spain.}
\footnotetext[16]{Supported by Bundesministerium f\"ur Bildung
und Forschung, Germany.}
\footnotetext[17]{Supported by the Direction des Sciences de la
Mati\`ere, C.E.A.}
\footnotetext[18]{Supported by the Austrian Ministry for Science and Transport.}
\footnotetext[19]{Now at SAP AG, 69185 Walldorf, Germany}
\footnotetext[20]{Now at Groupe d' Astroparticules de Montpellier, Universit\'e de Montpellier II, Montpellier, France.}
\footnotetext[21]{Now at BNP Paribas, 60325 Frankfurt am Mainz, Germany}
\footnotetext[22]{Supported by the US Department of Energy, grant DE-FG03-92ER40689.}
\footnotetext[23]{Now at IIHE, CP 230, Universit\'{e} Libre de Bruxelles, 1050 Bruxelles, Belgique}
\footnotetext[24]{Now at Dipartimento di Fisica e Tecnologie Relative, Universit\`a di Palermo, Palermo, Italy.}
\footnotetext[25]{Deceased.}
\footnotetext[26]{Now at SLAC, Stanford, CA 94309, U.S.A}
\footnotetext[27]{Now at CERN, 1211 Geneva 23, Switzerland}
\footnotetext[28]{Research Fellow of the Belgium FNRS}
\footnotetext[29]{Research Associate of the Belgium FNRS} 
\footnotetext[30]{Now at Liverpool University, Liverpool L69 7ZE, United Kingdom} 
\footnotetext[31]{Supported by the  Interuniversity Attraction Pole P5/27} 
\footnotetext[32]{Now at Henryk Niewodnicznski Institute of Nuclear Physics, Cracow, Poland}
\footnotetext[33]{Supported by US National Science Foundation grant PHY-0854724.}
\footnotetext[34]{ Supported by the James Arthur fellowship.}
\setlength{\parskip}{\saveparskip}
\setlength{\textheight}{\savetextheight}
\setlength{\topmargin}{\savetopmargin}
\setlength{\textwidth}{\savetextwidth}
\setlength{\oddsidemargin}{\saveoddsidemargin}
\setlength{\topsep}{\savetopsep}
\normalsize
\newpage
\pagestyle{plain}
\setcounter{page}{1}